\documentclass[]{aastex631}

\usepackage{graphicx}
\usepackage{amsmath}
\usepackage{amssymb}
\usepackage{verbatim}
\usepackage{multirow}
\usepackage{bigdelim}
\usepackage{threeparttable}
\usepackage{float}
\usepackage{soul}

\newcommand{\usc}{Department of Physics \& Astronomy, University of South Carolina, Columbia, SC 29208, USA}
\newcommand{\geo}{Department of Biochemistry, Chemistry, \& Physics, Georgia Southern University, Statesboro, GA, 30458, USA}

\shorttitle{Dust in distant galaxies}
\shortauthors{Klimenko et al.}

\begin{document}
\title{Probing Dust Composition in Distant Galaxies with JWST Mid-IR Spectroscopy of Quasars with Foreground 2175 {\AA} Absorbers \textbf{I: Methodology}}

\author[0000-0001-6730-2917]{Viacheslav V. Klimenko}
\affiliation{\usc}
\author{Varsha P. Kulkarni}
\affiliation{\usc}
\author{Monique C. Aller}
\affiliation{\geo}

\begin{abstract}
Interstellar dust plays a crucial role in gas cooling and molecule formation, influencing galaxy evolution. However, the composition and structure of dust in distant galaxies are still poorly understood. We have started a JWST MIRI MRS program investigating the dust features in gas-rich and dusty galaxies at redshifts $z<$1.2, with strong 2175~\AA\ bumps detected in absorption along the lines of sight to distant background quasars. 
{Here we describe our program strategy, and present MIRI MRS observations of IR dust features at $z=0.5-1.2$ in five quasar spectra that form the first part of our full sample. 
We identify artifacts in MIRI MRS data that affect the background in IFU cubes, and propose methods to reduce their effects.
We pay special attention to modeling the quasar mid-IR continuum, which shows significant variation depending on AGN morphology, redshift, and black hole mass. Dust in foreground galaxies produces significant absorption from the 10~$\mu$m silicate feature in all five quasar spectra. Compared with the average 10~$\mu$m silicate feature in the diffuse ISM of the Milky Way, we find differences in the absorption peak position, width of the features, and asymmetry of the profiles. A detailed study of these silicate features is presented in our next paper (Klimenko et al. 2026b). 
In two quasar spectra, we tentatively detect weak IR features near 3.0 and 3.4~$\mu$m. Their strengths are comparable to those seen in the Milky Way ISM, but follow-up observations are required to confirm these detections.}

\end{abstract}

\accepted{for publication in ApJ. DOI:~10.3847/1538-4357/ae3240}


\section{Introduction}
\label{sec:intro}

Dust makes up only $\sim$1\% of the interstellar medium (ISM) mass in
galaxies such as the Milky Way. Yet, it has a tremendous influence on the physics and chemistry of the ISM \citep[e.g.,][]{Draine2003}.
Dust shields gas from ultraviolet (UV) radiation, and provides catalytic surfaces for the formation of molecules such as H$_2$, thereby facilitating cooling in star-forming clouds. A robust understanding of the nature, formation and processing of dust is thus fundamental to understanding star formation and chemical enrichment in galaxies. Furthermore, dust can radically change the appearance of stars and galaxies, since it attenuates UV and visible light {(with a much weaker effect in the near-infrared (near-IR) and mid-IR)} and re-radiates it in the IR. Accurately correcting for these dust effects is crucial in order to infer properties that lie at the basis of many cosmological and astrophysical studies, such as the distances to standard candles, and the stellar properties of galaxies at known distances (e.g., luminosities and star formation rates). 

Unfortunately, our understanding of interstellar dust in distant galaxies is still very limited, and many extragalactic studies often assume extinction curves of distant galaxies to be similar to those of the Milky Way (MW), the Large or Small Magellanic Clouds (LMC/SMC), or local starburst galaxies \citep{Calzetti2000}. Such assumptions are too simplistic, especially since dust is expected to evolve in composition with time as a galaxy matures. Dust in a young galaxy is formed predominantly from core-collapse supernovae (SNe), while ejecta from asymptotic giant branch (AGB) stars contribute to dust in an older galaxy \citep[e.g.,][]{Galliano2008,Dwek2011,Slavin2020}. To test such basic expectations about dust evolution with cosmic time, and to improve the accuracies of dust corrections, it is vitally important to obtain direct measurements of dust properties in galaxies at cosmological distances. 

The primary ingredients of dust grains are believed to be carbonaceous materials, such as graphite or polycyclic aromatic hydrocarbons (PAHs), and silicates containing refractory elements, such as Mg and Fe \citep[see, e.g.][]{Li2001,Hensley2023}. 
{Carbonaceous and silicate dust grains are efficiently produced and expelled in asymptotic giant branch (AGB) stellar winds \citep[e.g.][]{Ferrarotti2006, Schneider2014} and in the ejecta of core-collapse SNe \citep[][]{Bocchio2016, Chastenet2022}}.
A large fraction of the ejected interstellar dust subsequently grows and is processed in the ISM itself. 
The composition of dust grains and their size distribution affect the wavelength-dependence of their extinction properties. As a
galaxy evolves, the grain composition is expected to evolve, with AGB stars contributing increasingly at later epochs. Furthermore, the grain size distribution evolves due to processes such as accretion, coagulation, sputtering, and grain-grain collisions. These changes are predicted to  have a profound effect on the extinction curve of the dust, with a steeper UV slope in a more evolved galaxy compared to that in a young galaxy \citep[e.g.][]{Asano2014}. 

However, determining the detailed properties of interstellar dust grains is difficult for distant galaxies, which are often too faint to detect in stellar or nebular emission, unless they are strongly star-forming, e.g., following a merger or starburst. A powerful technique for studying the dust in more normal, undisturbed galaxies, with modest levels of star formation, is to observe them in absorption, rather than emission, against the light of bright background sources, such as quasars with reasonably well-determined intrinsic spectra (i.e., using absorption features imprinted on the background quasar’s spectrum by the foreground galaxy’s ISM). Since this absorption-line technique depends only on the amount of ISM along the sight line, it is independent of the absorbing galaxy’s redshift or current level of star formation. This technique has been used widely in the UV and at visible wavelengths to study the gaseous component of the ISM and also the circumgalactic medium (CGM). Indeed, the technique has been well-utilized by UV and optical spectrographs on ground-based telescopes, as well as the Hubble Space Telescope (HST), for numerous quasars 
\citep[see, e.g.,][and references therein]{Bahcall1969, Bahcall1993, Petitjean1998b, Tumlinson2013, Tumlinson2017}. Such optical and UV spectral data can also be used to constrain the overall dust extinction in the galaxy sightline, as well as carbonaceous dust grain properties. A similar approach in the IR can probe absorption features of silicate dust grains, as well as possibly different species of carbonaceous dust grains, and thus expand the study of the dust phase in the ISM and CGM of distant galaxies. Past IR observatories such as the Infrared Space Observatory (ISO) and the Spitzer Space Telescope (SST) provided opportunities to study dust in some galaxies 
\citep[e.g.,][]{Soifer2008,Kulkarni2007,Kulkarni2011,Aller2012,Aller2014}. The far higher IR sensitivity of the James Webb Space Telescope (JWST) now presents a perfect opportunity to extend the ISM and CGM studies to the dust phase of larger and more diverse samples of {\em normal} distant galaxies.

The strongest dust spectral features that can be searched for in the UV and IR are the 2175 \AA\ bump, believed to arise from carbonaceous grains, and the 10 and 18~$\mu$m silicate absorption features that arise from the Si-O stretching and the O-Si-O bending modes \citep[e.g.,][]{Draine2003}. Most past studies of dust in distant galaxies in absorption focused on the 2175 \AA\ bump in UV/optical spectra. A number of studies have searched for the 2175 \AA\ bump from foreground galaxies in quasar spectra, and have together revealed a few hundred 2175 \AA\ absorbers from $\sim$100,000 quasars from the Sloan Digital Sky Survey (SDSS) \citep[e.g.,][]{Junkkarinen2004, Srianand2008, Noterdaeme2009, Kulkarni2011, Ledoux2015, Ma2015, Ranjan2018, Ranjan2020}. 
Spectroscopic studies of gamma-ray burst (GRB) afterglows have also revealed the detection of the 2175 \AA\ bump in many sightlines \citep[e.g.,][]{Liang2009, Zafar2012, Zafar2018}.These studies suggest differences in dust in the distant Universe (e.g., a weaker 2175 \AA\ bump than the Milky Way) compared to dust in the MW, SMC, or LMC.

Searches for silicate absorption are also extremely important, especially since $\sim$70\% of the core mass of dust grains is expected to arise from silicates \citep[e.g.,][]{Greenberg1999, Zubko2004}. With this in mind, members of our group pioneered observations of silicates in distant galaxies using background quasars. \citet{Kulkarni2007} reported the first detection of the 10 $\mu$m silicate absorption feature in a damped Lyman-alpha (DLA) absorber at $z=0.52$ using the Infrared Spectrograph (IRS) onboard the Spitzer Space Telescope. Follow-up studies by our group also succeeded in detecting silicate absorption in sight lines to background quasars passing through other foreground
galaxies at z$<$1.4, spanning a range of galaxy morphologies and impact parameters\footnote{Impact parameter is used to describe the projected distance between the quasar line of sight and a foreground galaxy.} \citep[e.g.,][]{Kulkarni2011,Kulkarni2016,Aller2012,Aller2014}. These studies found a variety of shapes and central wavelengths for the 10~$\mu$m absorption feature in galaxies at redshifts of 0.4-1.4, suggesting that silicate grains in these younger galaxies differ in composition and/or structure from grains in the MW ISM. In a few quasar absorbers, we reported the possibility of the silicates being crystalline  \citep[e.g.,][]{Aller2012,Aller2014,Kulkarni2016}. The existence of crystalline silicates in the ISM of galaxies, if confirmed, would be surprising, given that the silicates in the MW diffuse ISM are found to be largely ($>$95\%) amorphous \citep{Kemper2004,Li2007}. Despite these successes, past studies were limited by the low spectral resolution of the Spitzer IRS. Higher resolution spectroscopy is essential to more definitively determine whether the composition and structure of dust grains in distant galaxies indeed differs from that in the MW and nearby galaxies. 

The medium-resolution spectrograph (MRS) in the Mid-Infrared Instrument (MIRI) onboard the James Webb Space Telescope (JWST) offers tremendous advantages compared to the Spitzer IRS for studying dust grain composition and structure in distant galaxies. With a light-collecting area $>45$~times larger than Spitzer, the JWST offers much-needed higher sensitivity for achieving an adequate signal-to-noise ratio (SNR) in the spectra of foreground-dust-obscured, distant quasars. Furthermore, with a spectral resolving power $R \sim 1500-3000$, the MRS is much better equipped to resolve structure within dust spectral features compared to the Spitzer IRS (whose low-resolution mode with $R \sim 150$ was used in past studies of the silicate absorption from distant galaxies in the spectra of background quasars). Moreover, the MRS is an integral field spectrograph (IFS), thereby allowing spectroscopy of not just the background quasar, but also in other nearby sources (including the foreground absorbing galaxy, the quasar host galaxy, and any other field galaxies that may or may not be associated with the quasar or the absorber, Klimenko et al., in prep.). 

This paper is the first in a series from our study of dust infrared features detected with JWST in intervening dust-rich absorbers situated along quasar sightlines.
In this paper, we present our JWST MIRI MRS observations, and measure the 10~$\mu$m silicate absorption features in the spectra of five quasars. 
Section~\ref{sec:data} describes the sample selection, observations, and data reduction. Section~\ref{sec:continuum_model} describes the continuum fitting of the background quasars. In section~\ref{sec:results}, we present dust features detected in quasar spectra, and we summarize our results in Section~\ref{sec:conclusions}.

\section{Observations and data reduction}
\label{sec:data}

\subsection{Sample selection}
\label{sec:sample}

We aim to examine the composition and structure of dust grains in distant galaxies by primarily focusing our study on the strongest dust features in the UV range (2175~\AA\ bump) and in the IR range (10~$\mu$m silicate absorption). For this purpose, we selected a sample of absorbers detected in the spectra of quasars with available UV/optical spectra that clearly show strong 2175\,\AA\ dust features in the absorbing systems. 
A number of such 2175\,\AA\ bump systems have been detected in UV and/or optical spectra of background quasars obtained with the HST or SDSS \citep[e.g.,][]{Junkkarinen2004, Zhou2010, Jiang2011, Ma2017}.

Our sample targets were selected to satisfy the following criteria: 

\noindent
(1) Strong Mg~{\sc ii} absorption lines ($W^{\rm rest}_{\rm MgII2796}>2$~\AA) are detected in archival UV/optical spectra. Strong Mg~{\sc ii} absorbers at $z\sim1$ commonly trace dusty, metal-rich gas from the absorbing galaxies at small impact parameters of $\le30$~kpc \citep[e.g.,][]{Lundgren2021, Joshi2025}.
\\ 
(2) The absorber shows a strong 2175 \AA\ bump (with a reported bump strength, $A_{2175}$ $>$ 0.6 mag) in the spectrum of the background quasar.
\\
(3) The background quasar shows detectable reddening (0.1-0.3 mag), as determined from extinction curve fits of the SDSS/HST spectra published in the literature,  confirming the presence of a dusty foreground absorber. \\
(4) The absorber redshift is $z_{\rm abs} < 1.2$, so that the broad, redshifted 10~$\mu$m silicate feature and the nearby continuum could be covered in the MIRI MRS spectra.  \\
(5) The background quasar is reasonably bright in the mid-IR, as judged from the WISE W3 or W4 band photometry (with $8 < W3 <11$ mag and $5 < W4 < 8$ mag in most cases), so as to obtain adequate SNR in the continuum near the 10 $\mu$m feature (SNR $\sim$ 20-30 per spectral pixel) in a reasonable amount of integration time ($\sim$160-1600 seconds) with MIRI MRS. The targeted SNR was chosen with the goals of detecting the 10 $\mu$m feature at $>$20$\sigma$ significance, and reliably 
detecting substructures (if present) within the 10 $\mu$m feature at $>$5$\sigma$ significance.

Here we present observations of five targets, which were observed with JWST MIRI MRS in Cycle~1, and form the first part of our full sample. 
Targets properties are summarized in 
Table~\ref{tab:intro}. 

\setlength{\tabcolsep}{2pt}
\begin{table*}
\begin{center}
\caption{The properties of the sample of quasars with foreground absorption systems. The W3 magnitudes are the profile-fit values (from the AllWISEsurvey, \citealt{Cutri2013}). The values of the {absorption redshift},  strength of 2175~\AA\ bump ($A_{2175}$) and dust extinction  ($A_{V}$) are the published values from the listed references. The rest equivalent widths of Mg~{\sc ii} $\lambda$~2796 absorption line were measured from SDSS spectra. The targets were judged to be dusty based on the 2175~\AA\ bump strengths, and $A_{V}$ values where available.
}
\label{tab:intro}
\begin{tabular}{|l|c|c|c|c|c|c|c|}
\hline
  Quasar & W3 (12~$\mu$m) & $z_{\rm qso}$ & $z_{\rm abs}$ & $A_{2175}$ & $W^{\rm rest}_{\rm MgII2796}$ & $A_V$  &   References \\
   &  mag & &        &  mag & \AA        & mag  & \\
\hline
 AO0235+164$^\star$ & $7.95\pm0.02$ &$0.940$ & $0.524$ & $\sim3$ & $2.42\pm0.20$ & $\sim0.5$ & (a), (b) \\ 
 J0900+0214 & $10.86\pm0.11$ & $1.992$ & $1.051$ &  $2.06\pm0.60$ & $3.95\pm0.24$ & ..            &  (c)\\ 
 J0901+2044 & $9.63\pm0.05$ & $2.099$ & $1.019$ & $0.64^{+0.34}_{-0.24}$ & $2.04\pm0.05$ & $0.22^{+0.40}_{-0.20}$ &  (d)\\ 
 J1007+2853 & $8.94\pm0.03$ & $1.047$ & $0.884$ & $3.60\pm0.20$ & $3.34\pm0.17$ & $1.08^{+0.11}_{-0.11}$ &  (e) \\ 
 J1017+4749 &  $10.69\pm0.10$ & $1.219$ & $1.118$ & $1.40\pm0.35$ & $2.78\pm0.13$ & ..    &   (c) \\ 
\hline
\end{tabular}
 \begin{tablenotes}
      \small 
     \item{References:} (a) \cite{Junkkarinen2004}; (b) \cite{Kulkarni2011}; (c) \cite{Jiang2011}; (d) \cite{Ma2017}; (e) \cite{Zhou2010}.
     \item{Notes:}   
     $^{\star}$AO0235+164 has previously been observed with Spitzer/IRS. Detection of 10~$\mu$m silicate feature confirmed the presence of silicate-rich dust in the intervening absorber \citep[see][]{Kulkarni2007}. 
    \end{tablenotes}
\end{center}
\end{table*}

\subsection{Data Acquisition}
The data reported in this study constitute a part of the data obtained in our Cycle 1 JWST observing program (PID 2155, PI V. P. Kulkarni). The observations were carried out using the MIRI MRS \textbf{between} December 2022 \textbf{and} April 2023. In order to cover the features of interest and sufficient continuum, the MRS observations were obtained in channels 1, 2, 3, and 4 with sub-bands A, B, and C, thereby obtaining continuous coverage at 4.9-28.9~$\mu$m (observed-frame). 
Additionally, dedicated background exposures offset by $\sim$10\arcsec~from the science targets were obtained for each target, with the goal of better determining the background (which is strong in the mid-infrared due to contributions from the zodiacal light, as well as thermal self-emission from the telescope primary mirror, sun-shield, and other spacecraft components). 
The background observations were especially necessary because the small field of view (FOV) of the MRS could not contain sufficiently broad background regions around the background quasar, the foreground absorbing galaxy, and the quasar host galaxy, which together fill most of the {MIRI MRS IFU} frame\footnote{The field of view of MIRI MRS is $3.2\arcsec \times 3.7\arcsec$ for Channel~1 and $6.6\arcsec \times 7.7\arcsec$ for Channel~4, corresponding to projected physical scales of approximately 25~kpc and 50~kpc, respectively, at $z = 1$. It covers the quasar and likely includes the absorber's host galaxy, if separated by $\leq30$~kpc. Note that the MIRI MRS IFU image is also smoothed by convolution with the MRS PSF (FWHM = 0.5-1\arcsec).}. 
To improve spatial sampling and correct data for bad pixels, we used a 4-point dither pattern optimized for extended sources for the target exposures, and a 2-point dither pattern for the background exposures. For AO0235+164 only, the background was observed with a 4-point dither pattern to improve background quality, since the field of AO0235+164 is known to be rich in objects \citep{Chun2006}. Table~\ref{tab:obslog} summarizes our observations.

\setlength{\tabcolsep}{2pt}
\begin{table*}
\begin{center}
\caption{JWST MIRI MRS observing log. The `Settings' column lists the number of groups per integration ($N_{\rm grp}$), the number of integrations per exposure ($N_{\rm int}$), and the number of dithers. $T_{\rm exp}$ represents the total exposure time in seconds. The final four columns provide the average signal-to-noise ratio (SNR) in the MIRI MRS spectra for channels 1 through 4.} 
\label{tab:obslog}
\begin{tabular}{|l|c|c|c|c|c|c|c|c|c|c|c|}
\hline
  Quasar & RA        & DEC        & Date & \multicolumn{3}{c}{Settings} & $T_{\rm exp}$ & \multicolumn{4}{c|}{SNR} \\
         &                  (J2000.0) & (J2000.0) &       & $N_{\rm grp}$ & $N_{\rm int}$ & $N_{\rm dith}$ &  (s) &  Ch1 & Ch2 &Ch3 & Ch4 \\
\hline
 AO0235+164 & 02:38:38.9301 & 16:36:59.28& 2023-01-20 & 15 & 1 & 4 & 166 & 26 & 36 & 62 & 12\\ 
 AO0235+164(BKGR) & 02:38:38.9301 & 16:36:49.28 & 2023-01-20 & 15 & 1 & 4 & 166 & .. & .. & .. & ..\\ 
 J0900+0214 & 09:00:16.6775 & 02:14:45.85 & 2023-04-03 & 46-56-74 & 2 & 4 & 1032-1254-1653 & 12 & 24 & 33 & 7 \\ 
 J0900+0214(BKGR) & 09:00:16.6775 & 02:14:35.85 & 2023-04-03 & 46-56-74 & 1 & 2&   255-310-410   & .. & .. & .. & ..\\ 
 J0901+2044 & 09:01:22.6797 & 20:44:46.55 & 2023-04-17 & 30 & 1 & 4 & 333 & 15 & 30 & 40 & 12\\ 
 J0901+2044(BKGR) & 09:01:22.6797 & 20:44:36.55 & 2023-04-17 & 30 & 1 & 2 & 166 &   .. & .. & .. & ..\\ 
 J1007+2853 &10:07:13.6744 & 28:53:48.40 & 2022-12-29 & 20 & 1 & 4 & 222 & 27 & 45 & 58 & 31\\ 
 J1007+2853(BKGR) & 10:07:12.9130 & 28:53:38.40 & 2022-12-29 & 20 & 1 & 2 & 111  & .. & .. & .. & ..\\ 
 J1017+4749 &10:17:51.1503 & 47:49:40.04 & 2022-12-29 & 60 & 1 & 4 & 666 & 30 & 38 & 50 & 20\\ 
 J1017+4749(BKGR) & 10:17:52.1434 & 47:49:40.04 & 2022-12-29 & 60 & 1 & 2 & 333  & .. & .. & .. & ..\\ 
\hline
\end{tabular}
\begin{tablenotes}
      \small
     \item{ Notes: For J0900+0214, the science target and background fields were observed with a different number of groups depending on the MIRI MRS band: 46 for Short, 56 for Medium, and 74 for Long. For other observations, the number of groups remained consistent across different bands.}
    \end{tablenotes}
\end{center}
\end{table*}

\subsection{Data Reduction}
The data were reduced using a custom modification of the JWST MRS calibration pipeline, version 1.14.0, along with calibration data files  ``jwst\_1185.pmap''. We applied the standard JWST calibration pipeline stages (\textit{Detector1Pipeline}, \textit{Spec2Pipeline}, and \textit{Spec3Pipeline}), with specific modifications in select stages to improve data quality. Additionally, background subtraction was performed at Stage 2 for the detector rate images, effectively reducing potential detector artifacts. 

We used the recommended steps from the \textit{`Detector1Pipeline'} stage for basic detector-level corrections and slope calculations in MIRI data, including correction for saturated pixels, first and last groups, reset anomaly correction, linearity correction,  reset switch charge decay (RSCD) correction, dark subtraction, reference pixels correction, jump detection, and ramp fit. 
However, the \textit{RSCD}, \textit{Jump Detection}, and \textit{Ramp Fit} steps were modified to 
incorporate recent corrections and updated algorithms, as described below. 

The RSCD effect creates nonlinearity in pixel slopes at the beginning of the integrations, due to a residual charge left over after the detector reset. By default, the RSCD correction drops several first groups for all second and higher integrations, assuming that the number of resets before the first integration is sufficient for the effect to have decayed from the previous exposure.
We found, however, that the nonlinearity persisted in the first integration, even after the default linearity correction was applied \citep[see also, e.g.,][]{Grant2023}.
To mitigate this effect, we modified the \textit{RSCD} step by dropping \textbf{a} specific number of first groups ($N_{\rm first}>1$) at the beginning of each integration.
The exact number of dropped groups depends on the total number of groups per integration ($N_{\rm grp}$): $N_{\rm first}=2$ for  $N_{\rm grp}< 20$; 6 for $20\le N_{\rm grp}<40$; and 12 for $N_{\rm grp}\ge 40$.

The original pipeline processes the \textit{Jump Detection} and \textit{Ramp Fit} steps sequentially, despite their interdependence, to optimize computational time. 
The procedure uses a criterion for masking groups affected by cosmic rays (jumps) based on the comparison between a local two-point slope (between two groups) and the overall average slope. This slope is initially unknown and is derived iteratively, based on the slope from unmasked groups. This original  procedure fails for pixels with a non-linear slope; the linearity correction and dark subtraction does not adequately account for non-linearity effects in all cases. 

We, thus, implemented a revision to the code using a custom algorithm for detecting cosmic ray events and calculating the slope based on a likelihood-based formalism, see \cite{Brandt2024}. This approach enables hypothesis testing for cosmic ray hits using the entire ramp while performing both operations (jump detection and slope fitting) simultaneously and consistently.
We found that this modification requires a comparable amount of computational time, and yields a slightly improved fit to the data. 
For J0900$+$0214, which has two integrations, the pixel slope images (rate files) were calculated separately for each integration. The rate files were then combined using the median method with additional corrections for cosmic ray showers (see the description of the procedure for the background exposures below). 

The final products of the \textit{`Detector1Pipeline'} (rate files) exhibit artifacts, wherein some pixels in the rate image are significantly brighter than their surroundings. These artifacts exhibit characteristic cross-like or point-like structures. Bright pixels were identified as 3$\sigma$ outliers within a radius of 3 pixels around a given pixel. To confirm that these pixels are detector artifacts, rather than unresolved background sources or spectral features, we masked only the `bright' pixels that were present in both the target and background exposure images taken under the same detector settings. Consequently, we identified approximately 3,000 `bright' pixels in each exposure, and corrected their values using linear interpolation along the dispersion axis.

Next, we applied the steps from the \textit{'Spec2Pipeline'} to conduct additional instrumental corrections and calibrations for images of both the target fields and the background fields. This process included applying \textit{AssignWCS}, \textit{Flat field correction}, \textit{Source identification}, \textit{Stray light correction}, \textit{Fringe flat correction}, and \textit{Residual fringe correction} to all exposures for both the target and background fields. We found that the background subtraction procedure becomes more efficient when applied to detector images, as opposed to subtracting the background in the cube data.
We applied it after the instrumental corrections ({flat field, stray light and fringe corrections}), which reduces instrumental effects that could complicate finding the optimal background model solution. 

\subsection{Additional Reduction Steps}

The background exposure rate images show variations in brightness due to weak cosmic ray hits, 
cosmic-ray showers (residual effects of energetic cosmic-ray hits, causing a slight increase in the brightness of nearby pixels due to charge inter-pixel migration), and instrumental artifacts, such as scattered light. 
We pay particular attention to correcting the data for these distortions, as their amplitudes and spatial scales in the detector image plane are similar to those of the weak, broad dust absorption features we are searching for in our data.

If we assume that the background fields do not contain unresolved sources, then the brightness of a pixel mainly depends on its detector coordinates rather than sky coordinates, allowing us to compare the brightness across dithered background images.\footnote{Additionally, we examined SDSS images of the background fields and the MIRI detector rate images for any signal from unresolved sources (which would appear as a trace along the dispersion axis for continuum sources in MIRI detector rate images), but we did not detect any.} 
Simple averaging pixel-by-pixel across the dithered images only reduces the distortions by a factor corresponding to the number of dithers, and does not fully remove them. To better address the effect, we developed a custom procedure,  described below in Appendix~\ref{app:A:Background-model}.

Background models are then subtracted from the target exposures. It is worth noting that CR showers (due to CR hits of pixels in target images) remain in the target exposures.
Additionally, target exposures include emission from our quasars, and may show emission of the absorbers' host galaxies, which appear as traces along the vertical (dispersion) axis of the detector image. Although CR showers impact different regions in dithered images, the dithering process also shifts the position of the quasar and absorber host galaxy traces.
Thus, our custom procedure for removing CR showers and instrumental artifacts in background exposures would also affect the pixel intensity in the quasar and host galaxy traces. We note, however, that this procedure remains effective for exposures that have with multiple integrations (available for J0900+0214) since the quasar trace positions remain fixed within each integration. 
For the remaining targets (which have a single integration per exposure), we developed an alternative custom procedure that allows us to correct for CR showers in the cube data (see details in Appendix~\ref{app:A:res_bkgr_corr}).

We also report an effect that contaminated MIRI MRS data for two of our quasars, appearing as scattered light stripes across the detector image, which, to our knowledge, has not been previously described in the literature. 
This artifact remains visible over time, thus impacting data with different dithers and bands. Appendix~\ref{app:A:artifact} presents an example of the artifact, and details the corrections applied to the data.

Next, we applied the \textit{Flux Calibration} step, and processed data through the \textit{`Spec3Pipeline'}.
The cubes were created separately for each of the 4 dithered images, the 4 channels, and the 3 bands (short, medium, and long), resulting in a total of 48 cubes per quasar.
MIRI MRS cube data are known to be affected by residual oscillations (``wiggles'') \citep{Gasman2023}. We applied a correction using the procedure described by \cite{Perna2023}. This improves the spectra of individual spaxels (which is essential when studying absorber host galaxies; Klimenko et al., in prep.), but it has little effect on the combined spectrum extracted within the quasar aperture\footnote{The combined quasar spectrum extracted from a large aperture plays the role of a reference model for detecting oscillation signals. Subtracting the wiggles from individual spaxels makes them more similar to the reference, resulting in minimal changes to the combined spectrum.}. We found the results for the combined spectrum to be similar to those obtained by applying the pipeline procedure for fringe corrections in the 1-D spectrum.\footnote{The procedure \texttt{fit\_residual\_fringes\_1d} is from \url{https://github.com/spacetelescope/jwst/blob/main/jwst/residual_fringe/utils.py}.}   

Finally, we extracted all of the quasar spectra (1A to 4C, and 4 dithering positions)  using an aperture with a radius equal to the 2$\sigma$ width of the point spread function (PSF), which increases with wavelength \citep[see][]{Argyriou2023B}.
This aperture does not fully capture the wings of the PSF; however, we find it to be optimal for achieving the highest SNR in the quasar spectra, as background variations are more pronounced in spaxels at larger radii. Spectra with different dither parameters were combined using a sigma-clipping method.  The 12 spectral chunks were then merged into a single quasar spectrum using a multiplicative scaling method to ensure proper alignment of the overlapping regions from each chunk. The spectra were re-binned by factors of 4 and 2 for channels 1A to 2C and 3A to 4C, respectively, to match the size of the spectral resolution elements (a resolution of $\lambda/\Delta\lambda\sim2500$). 

\begin{figure*}
\begin{center}
    \includegraphics[width=1\textwidth]{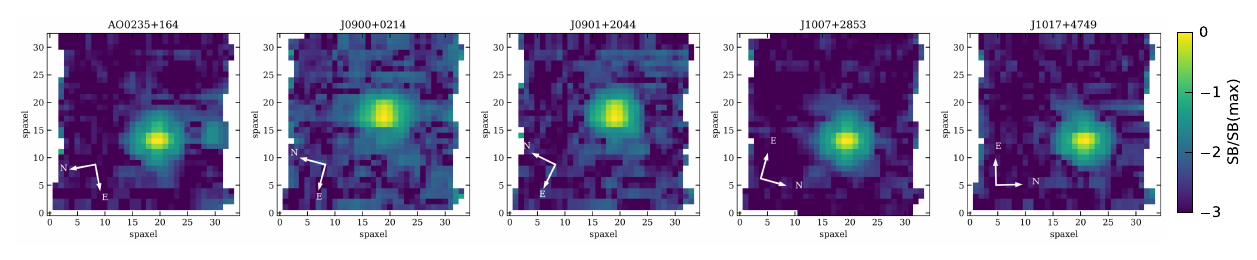}
    \includegraphics[width=1\textwidth]{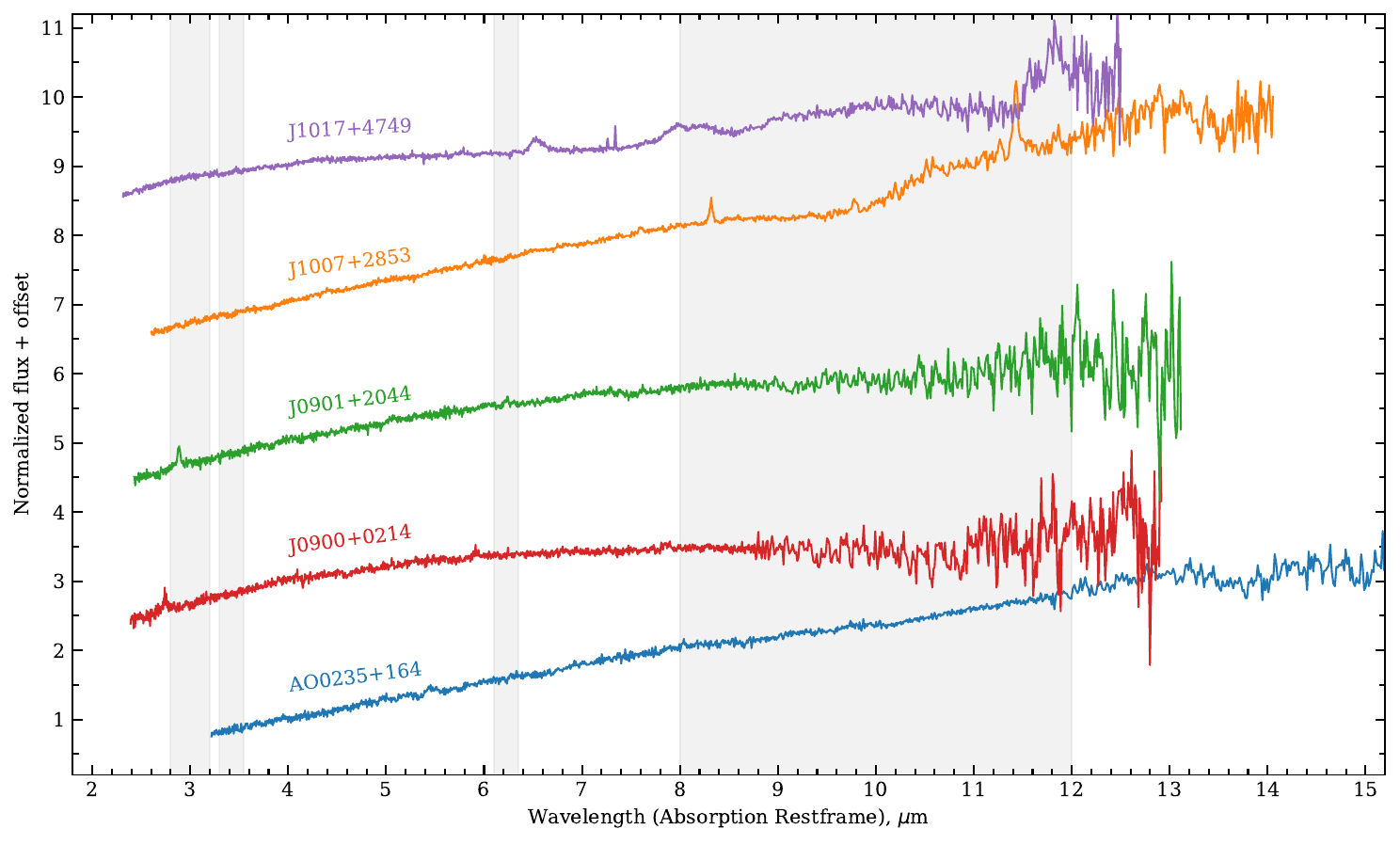}
    \caption{{\it Top:} Background-subtracted MIRI MRS images for our quasars. The images are shown for channel 3A, and presented in the IFU-aligned coordinate system. The color gradient represents the logarithm of the surface brightness (SB), averaged along the dispersion axis and normalized to unity at the brightest spaxel. The residual emission surrounding the quasars originates from the wings of the MIRI PSF, and has a characteristic hexagonal shape. The variation in the background around quasars is usually less than a few percent. For AO0235+164, we detect emission from two nearby galaxies, visible as bright spots to the east and south of the quasar. {\it Bottom:} MIRI MRS spectra (rebinned by a factor of 4, corresponding to a spectral resolution of $\sim600$) for the quasars, shown in the absorption system rest-frame. Spectra are normalized to unity near 4~$\mu$m and vertically offset by 2 units. The increasing uncertainties in the spectra at large wavelengths are due to the rising thermal noise of the detector in channel 4C. The vertical shaded regions indicate the wavelength ranges of 3.0~$\mu$m (H$_2$O), 3.4~$\mu$m (C–H), 6.2~$\mu$m (C-C) and 10~$\mu$m (silicate) absorptions.
    }
    \label{fig:cube-images}
\end{center}
\end{figure*}

Fig.~\ref{fig:cube-images} shows the quasar images within the field of view of the MIRI MRS (in channel 3A) and quasar spectra. We note that the flux in the background around the quasars is consistent with a zero level within a few percent, confirming that the dust features studied in the next section are not affected by instrumental artifacts.

\section{Constructing Intrinsic Quasar Continua}
\label{sec:continuum_model}

The MIRI MRS spectra span the wavelength range of 3 to 14~$\mu$m in the absorption rest frame for sources at $z\sim1$. In our sightlines, we could thus detect a broad $\sim$10~$\mu$m silicate feature (Si-O stretching mode), as well as weak, narrow absorptions from water ice at 3.0~$\mu$m, and features at 3.4~$\mu$m and 6.2~$\mu$m corresponding to C-H aliphatic and C=C olefenic/aromatic stretching modes in hydrocarbons, respectively. These features have been detected in MIRI MRS spectra of Milky Way stars probing the diffuse ISM ($A_V < 3$) \citep[see][]{Decleir2025}.

The strongest feature around 10~$\mu$m is clearly visible in two quasars, J1007$+$2853 and J1017$+$4749 (see Fig.~\ref{fig:cube-images}), and it is less prominent in the other three quasars (but it is also detected after a careful continuum normalization). 
Below we describe our procedure for reconstructing the intrinsic quasar continuum for the wavelength range of the 10~$\mu$m silicate feature. For the narrow dust features, we fit the continuum locally using a polynomial function.

Depending on the morphology of the active galactic nucleus (AGN or quasar), the mid-IR (MIR) AGN spectrum may include contributions from synchrotron emission, the accretion disk, the broad-line region, and the dust torus, as well as emission from the host galaxy stellar population and star formation (e.g., \citealt{Malmrose2011,Raiteri2014,Shi2014}). For example, some AGN spectra exhibit AGN 10~$\mu$m silicate emission or absorption and polycyclic aromatic hydrocarbon (PAH) emission at e.g., 11.3~$\mu$m \citep{Shi2014}. Furthermore, the AGN redshift may impact the shape of the spectral energy distribution (SED); \citet{Xu2015} illustrate that the relative contributions of the stellar photospheric component, star formation component, and AGN emission for Type-1 AGN vary such that at higher (z$\sim$1.5) redshifts the AGN dominates the average SED emission in the 1-20~$\mu$m region, whereas at lower (z$\sim$0.5) redshifts the stellar contributions can be significant.  

When matching the shape of the observed MIRI MRS spectrum to candidate continuum normalizations constructed over a broader spectral range, we generally rely on spectral data blueward of the absorber 10~$\mu$m feature. This is for two reasons. First, for our foreground absorbers ($0.5\leq z \leq 1.2$), the 10~$\mu$m feature is redshifted to $\gtrsim$15~$\mu$m in the observed frame, placing it in MIRI MRS channels 3-4. At wavelengths longer than the 10~$\mu$m feature, the detector thermal noise substantially reduces the SNR for faint sources. Second, depending on the grain mineralogy and the ratio of the 10-to-18~$\mu$m silicate absorption features, there may be silicate absorption redward of the peak 10~$\mu$m feature, which should not be considered in determining the continuum level.

Although representative continuum templates exist in the literature for differing AGN morphologies (e.g., \citealt{Elvis1994, Richards2006, Shi2014}), we find that these median/average SEDs do not suffice to normalize our AGN continua. This is because there can be a range of small spectral structural differences within the parent populations of AGN used in their construction, such as variations in emission component strengths or in the AGN silicate dust features. Therefore, for each source in our sample we independently determine the continuum shape using information on the AGN morphology, redshift, and black hole mass combined with Spitzer Infrared Spectrograph (IRS) spectra (5-37~$\mu$m, observed frame) for similar sources. Although the Spitzer IRS spectra are lower resolution than our JWST MIRI MRS spectra, they are more finely sampled than the photometric data used to construct AGN SEDs (e.g., \citealt{Richards2006}), and are adequate for determining the intrinsic AGN continuum shape.

\begin{figure*}
\begin{center}
\includegraphics[width=1\textwidth]{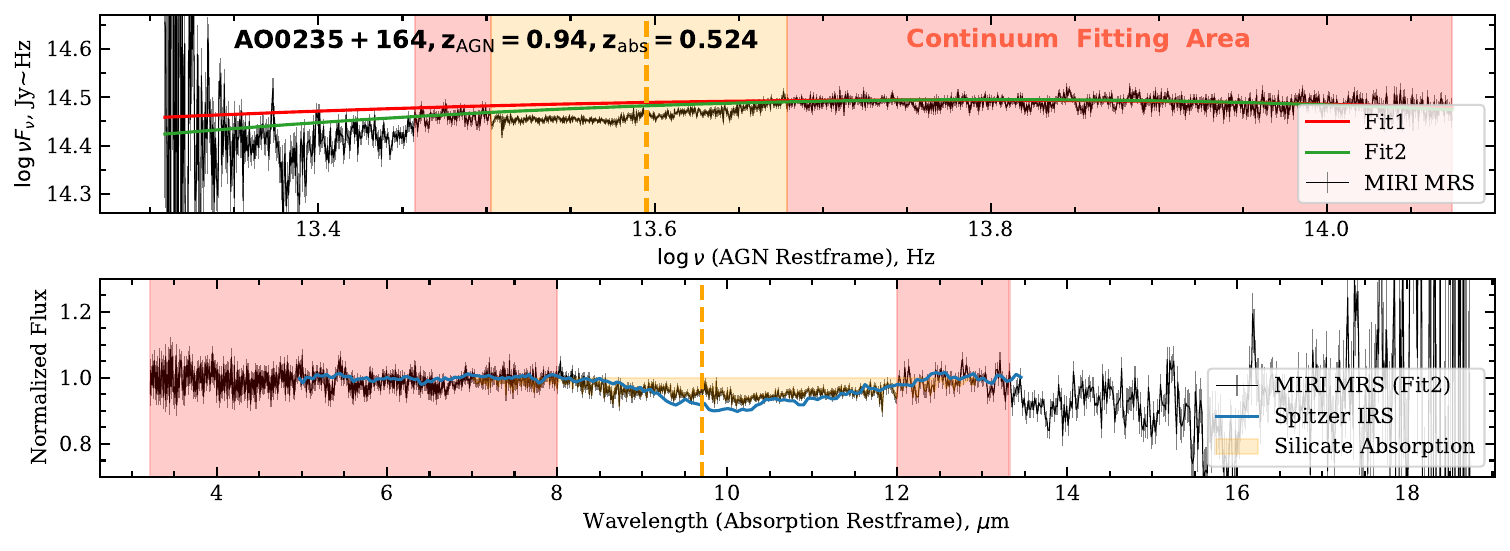}
\includegraphics[width=1\textwidth]{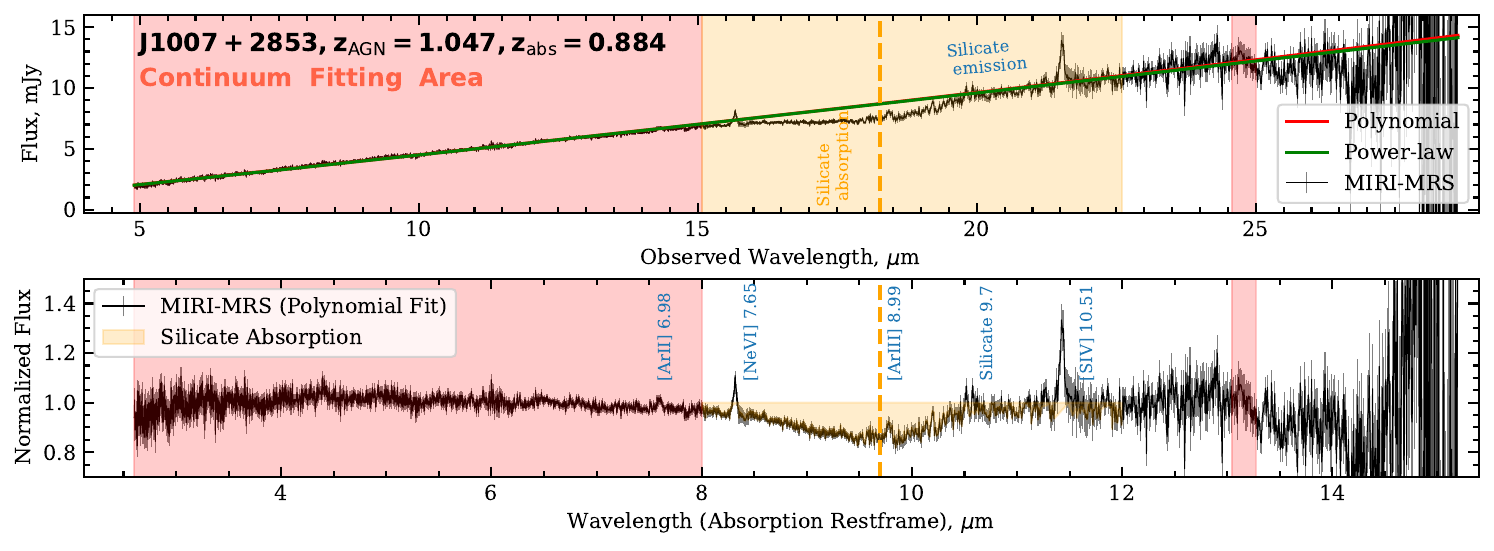}
\caption{Fit to the continuum and normalized spectrum for the blazar AO~0235$+$164 (upper two panels) and the radio-loud AGN~J1007+2853 (lower two panels). The black curve represents MIRI MRS data (binned by a factor of 2 for visual clarity). The red shaded regions indicate the portions of the spectrum used for continuum fitting. The orange shaded area marks the region with the foreground galaxy 10~$\mu$m silicate absorption feature. The orange vertical lines shows the position of 9.7~$\mu$m (the peak wavelength in Milky-Way-type silicate profiles) at the galaxy rest-frame. The green and red curves represent two fits for the continuum (see text for details). For AO~0235$+$164, the normalized spectrum from archival Spitzer IRS observations is shown in blue, closely matching the MIRI MRS data. For J1007+2853, the identified AGN emission lines are labeled in blue.}
\label{fig:j0235-j1007-continum}
\end{center}
\end{figure*}

\begin{figure*}
\begin{center}
        \includegraphics[width=\textwidth]{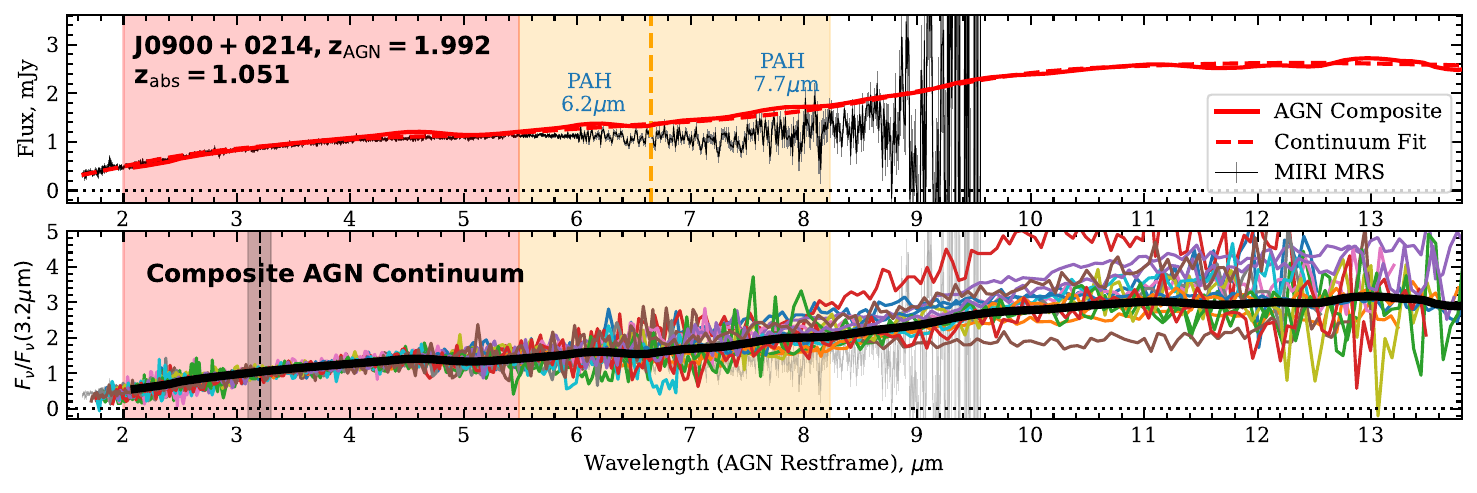}
        \includegraphics[width=\textwidth]{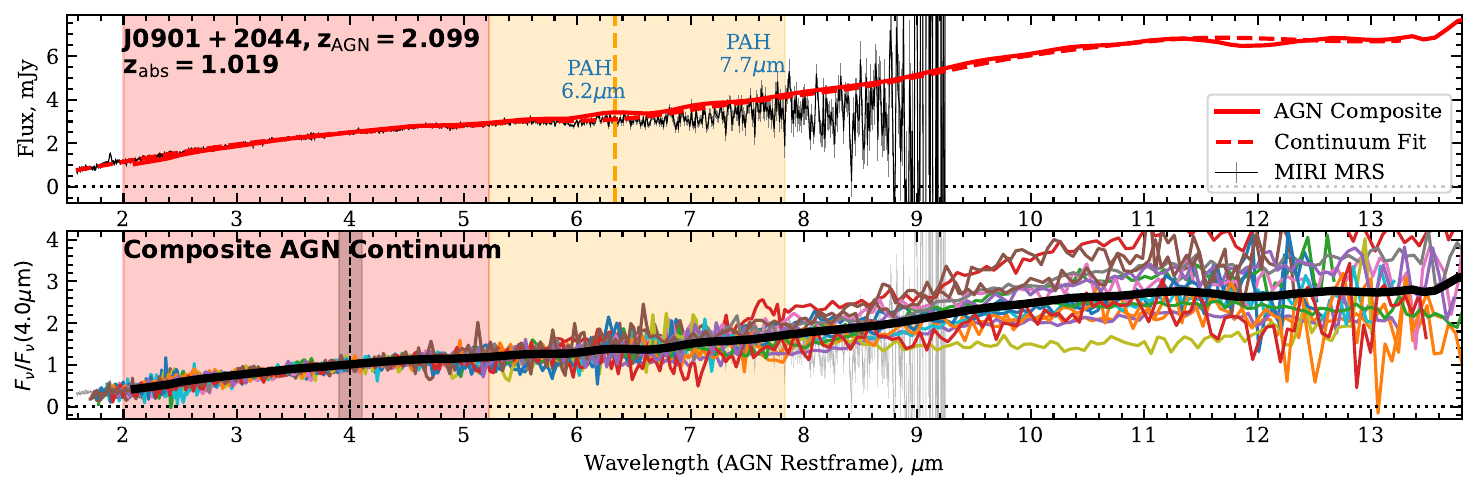}
        \includegraphics[width=\textwidth]{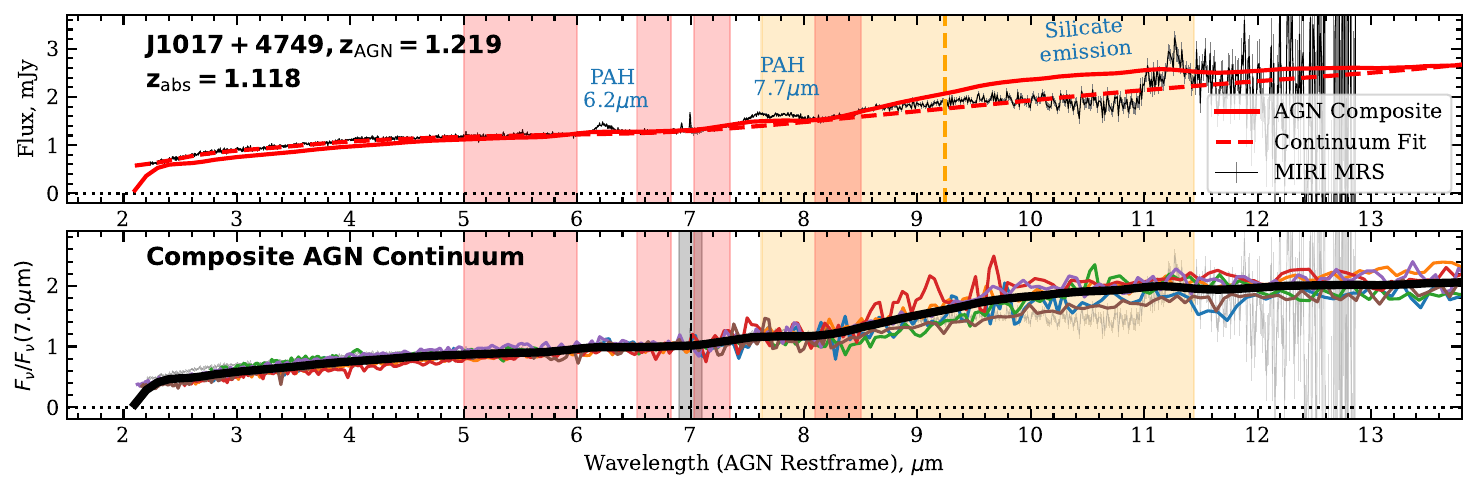}
        \caption{Fit to the AGN continuum for J0900$+$0214, J0901$+$2044, and J1017$+$4749. Top panels for each AGN show the MIRI MRS spectrum (in black, rebinned by a factor of 2) overlaid with the constructed composite AGN continuum (red solid line), along with a version that excludes emission features and is corrected for flux variations below 3-5~$\mu$m at the AGN rest frame (red dashed line). The red and orange shaded areas are the same as in Fig.\ref{fig:j0235-j1007-continum}.
        The bottom panels for each AGN show the Spitzer/IRS spectra of AGN with similar properties 
        used to construct the composite AGN continuum. The IRS spectra and the AGN composite are shown in different colors and in black, respectively. All spectra are normalized to unity at the wavelength ($\lambda_n$) marked by the gray-shaded vertical line. 
        }
        \label{fig:j0900-continuum}
\end{center}
\end{figure*}

\subsection{Blazar AO~0235+164 }
AO~0235+164 is a well-studied, temporally-variable, z=0.940 blazar that is alternately classified as a BL Lac or a flat spectrum radio quasar (FSRQ) depending on the observation epoch. A study by \citet{Raiteri2014} found that BL Lac SEDs are generally consistent with synchrotron emission from the jet, and possibly a contribution from the host galaxy, whereas FSRQs require synchrotron emission combined with contributions from the accretion disk, broad-line region, and dust torus. In the case of AO~0235+164, \citet{Raiteri2014} found that the MIR SED was dominated by synchrotron emission, but that notably the peak of the synchrotron emission in the MIR shifts to higher energies when the MIR peak flux is higher; thus, the SED shape determined at earlier epochs cannot be used for our normalization. During the epochs \citet{Raiteri2014} investigated, the peak occurred at $\log \nu_p ({\rm Hz}) \approx 13.0-13.5$, and \citet{Lister2019} give a value of  $\log \nu_p ({\rm Hz})= 13.0$, in the observed-frame. We follow the procedures of \citet{Massaro2004} and \citet{Raiteri2014} and model the jet synchrotron emission that dominates the SED as a log-parabola such that
\begin{equation}
    log[\nu F(\nu)]=log[\nu_p F(\nu_p)]-b(\log\nu-\log\nu_p)^2
\end{equation}
where $\nu_p$ is the synchrotron turnover peak frequency in Hz and $\nu_pF(\nu_p)$ is the peak emission. In order to avoid fitting over the region where the foreground absorber 10~$\mu$m silicate absorption feature is expected, and to minimize the contributions from the lower signal-to-noise, long-wavelength MIRI data, we perform two fits.  
In the first, we fit the SED only at $\log\nu>13.67$ in the AGN rest-frame ($\lambda < 8$~$\mu$m, absorber rest-frame), thereby avoiding the 10~$\mu$m absorber feature. In the second fit, we include the part of the SED at $13.45 \le \log \nu \le 13.5$ in the AGN rest frame ($12~\mu{\rm m} < \lambda < 13.3~\mu{\rm m}$, absorber rest-frame). This latter region may be influenced by silicate absorption, but provides a better fit to the spectrum at redder wavelengths. 
The best fits and continuum-fitting regions are shown in the top panels of Fig.~\ref{fig:j0235-j1007-continum}. Later, we consider both fits in the analysis of the silicate absorption line. The quasar spectrum normalized using these two fitting-functions appears similar near the 10$~\mu{\rm m}$ silicate absorption of the foreground galaxy. We find that for the epoch of our observations $\log \nu_p=13.80$ for the first fit (and $b=0.288)$, and $\log \nu_p=13.77$ for the second fit (and $b=0.167)$ in the quasar rest frame. The peak flux is about 3.3~mJy (or $\log \nu_pF(\nu_p)\simeq-11.5~{\rm erg~cm^{-2}~s^{-1}}$). We also analyzed the spectrum obtained with Spitzer IRS in 2006 (see data reduction details in \citealt{Kulkarni2007}), and found the best fit continuum model for $\log \nu_p=13.48$ and $\log \nu_pF(\nu_p)=-10.7~{\rm erg~cm^{-2}~s^{-1}}$.
When comparing the synchrotron peak wavelength with the flux at that location, we find that over four epochs of observation for this source, including the two epochs investigated by \citet{Raiteri2014}, our previous Spitzer IRS spectrum, and our new JWST MIRI spectrum, there is no correlation between these two properties, unlike the possible trend noted by \citet{Raiteri2014}. Observations of other blazars suggest these two properties may only be correlated in some flaring epochs \citep[e.g.,][]{Zuo2025}.

\subsection{AGN J1007+2853}
The AGN J1007+2853 is a z=1.047 core-dominant, non-Broad Absorption Line (BAL), radio-loud quasar with a supermassive black-hole (SMBH) mass of $\log M_{\bullet}(M_{\odot})=9.57 \pm 0.15$ \citep{Shen2011}. Although radio-loudness is expected to be more significant at longer wavelengths, the \citet{Elvis1994} templates exhibit small differences in the SED shape in the MIR for radio-loud (RL) and radio-quiet (RQ) quasars. Therefore, we identified similarly radio loud (radio-loudness $>$10, e.g. \citealt{Jiang2007}) sources with a core-dominant morphology in the \citet{Shen2011} sample of 105,783 quasars from the SDSS data release 7 (DR7). We further restricted our sample to non-BAL quasars within a redshift range of $0.50\leq z \leq 1.55$ and a SMBH mass range of $8.5 \leq \log M_{\bullet} \leq 10.5$, which did not have cataloged foreground absorbers. We then identified all such sources with a Spitzer IRS spectrum available in the Cornell Atlas of Spitzer/Infrared Spectrographic Sources \citep[CASSIS;][]{Lebouteiller2011} database of low- and high-resolution slit spectra. Of the $>$2,000 AGN that matched our search criteria, 9 sources had suitable data. The intrinsic spectral shape of these sources could be well-described by a linear (or low-order polynomial) or power-law fit, with small residuals. The MIRI MRS spectrum of J1007$+$2853 is likewise relatively smooth, but exhibits AGN's nebular emission lines of [Ar~{\sc ii}], [Ar~{\sc iii}], [Ne~{\sc VI}], and [S\,{\sc iv}], 
which were not present in the IRS template spectra. The 10~$\mu$m silicate feature associated with the $z_{\rm abs}=0.884$ foreground absorbing galaxy is strongly pronounced. Additionally, there is weak emission at 10~$\mu$m in the AGN rest-frame, which may be due to AGN silicate emission. In order to avoid fitting our MIRI MRS spectrum over the absorber or AGN silicate features, we restrict fitting to the ranges of $\lambda < 8(1+z_{\rm abs})~\mu{\rm m}$ and $12(1+z_{\rm AGN})~\mu{\rm m}<\lambda < 25~\mu{\rm m}$ in the observer rest-frame. To ensure that fitting only part of the spectrum would not significantly impact the extrapolated shape, we compared the effects of fitting the absorber-free template spectra over a spectral range including only the data at bluer wavelengths and extrapolating over the redder wavelengths with the full fits. We found small divergences for the power-law fits, but general consistency for the linear  fits. We, thus, adopt the low-order polynomial fit, $y(x)=-0.280 + 0.460 x + 0.0017 x^2$ for this source, wherein $y$ is the flux density and $x$ is the observed wavelength (in~$\mu$m).
The fit to the continuum for J1007+2853 is shown in lower panels of Fig.~\ref{fig:j0235-j1007-continum}.   
The red wing of the silicate absorption line may overlap with the blue wing of the AGN silicate emission line. The AGN silicate emission line appears weak in the observed spectrum, possibly due to suppression by the foreground silicate absorption. As the line superposition is unclear ambiguous, we consider two approaches by fitting the silicate absorption line with and without varying the strength of the AGN silicate emission component (see details in Paper II, Klimenko et al. 2026b).

\subsection{AGNs J0900+0214, J0901+2044, J1017+4749}
The remaining three sample AGN (J0900+0214, J0901+2044, J1017+4749) are non-BAL, radio-quiet AGN. In these sources, the intrinsic AGN continuum shape is more complex, and is not well-represented by a simple low-order polynomial or
power-law fit.  To determine the intrinsic SED shape, we selected objects from the \citet{Shen2011} catalog which were not detected in the FIRST radio survey \citep{White1997}, are not flagged as BALQSOs, and which have a redshift within a range of
 $z_{src}\pm0.5$ (with $z_{src}$=1.99, 2.10, 1.22 for J0900+0214, J0901+2044, and J1017+4749 respectively) and black hole mass $\log M_{\bullet,src}\pm1.0$~dex 
($\log M_{\bullet,src} (M_{\odot})=8.81\pm0.43$, $9.57\pm0.04$, $9.28\pm0.08$, respectively). We searched for all such sources in the CASSIS Spitzer IRS  database, eliminating those sources with known foreground absorbers or data quality issues. For the remaining candidate-template sources, the CASSIS spectral orders were stitched together, if needed, and the spectra were shifted into the AGN-rest-frame and normalized to unity at $\lambda_n$ (3.2~$\mu$m, 4.0~$\mu$m, and 7.0~$\mu$m, respectively), chosen to be within the fitting area and avoid spectral features for that source. We likewise normalized and redshifted our JWST spectra for comparison.  
The fitting regions were chosen to be $2-5.4~\mu$m, $2-5.2~\mu$m, and $5-8.5~\mu$m in the AGN rest frame, respectively. These regions were selected by excluding the foreground galaxy 10~$\mu$m silicate absorption feature and the  6.2~$\mu$m PAH and 10~$\mu$m silicate AGN emission features.
For J1017+4749, we observe a flattening of the spectrum below approximately 5~$\mu$m in the AGN rest frame, a feature seen in only a few of the templates.\footnote{A similar feature is also observed in the spectra of J0900+0214, J0901+2044 below 3~$\mu$m, although it is less prominent.} Thus, this region was excluded from the fit to focus on candidate-template sources whose spectra more closely match the continuum of J1017+4749 near the 10~$\mu$m absorption feature of interest. For illustrative purposes, we applied an additional correction for the flattening of the JWST spectra below 5~$\mu$m; the continuum was multiplied by a low-order polynomial function to account for this difference. 

Those candidate template sources with $\chi^2_{\rm red}<3-5$, when compared with the MIRI MRS spectrum, were used to construct a mean continuum template (composite), which was then smoothed with a Savitzky-Golay\footnote{The filter is used to smooth data while preserving the shape and
important details in the data (like emission or absorption lines) \cite[e.g.,][]{Schafer2011}. The SG filter is implemented in the SciPy library as {\sc scipy.signal.savgol\_filter}} filter  with a window length of 8 pixels. We utilized 16, 16, and 6 templates to calculate the mean template continuum for J0900+0214, J0901+2044, and J1017+4749, respectively. 
The fitting procedure is illustrated in Fig.~\ref{fig:j0900-continuum}.

For each AGN, the mean template (composite) exhibits a steeply increasing continuum, characterized by a power-law component, and includes the PAH 6.2~$\mu$m and 7.7~$\mu$m, and the 10~$\mu$m silicate, AGN emission features. Some of these emission lines (PAH lines for J0900+0214 and J0901+2044, and the silicate line for J1017+4749) overlap with the foreground galaxy silicate absorption feature, complicating the reconstruction of the continuum in these regions.
We, thus, consider two versions of the continuum: one represents the mean template, and the other consists of the continuum emission of the mean template derived by B-spline fitting the spectral regions not affected by AGN emission features. The second version is used to fit the silicate absorption of the foreground galaxy simultaneously with the AGN emission lines (see Paper II).

\section{Results}
\label{sec:results}

In this section, we present the profiles of the dust features in the targeted distant galaxies, and compare them with the typical profiles of dust features observed in the Milky Way ISM.   
A detailed analysis of the dust features - including fitting the 10~$\mu$m silicate profile with laboratory/astrophysical templates and theoretical models, and comparing its strength to other dust properties such as visual extinction, the strength of 2175~\AA{} bump, and extinction curve parameters   
in samples of our quasars and Milky Way sightlines
is presented in the Paper~II. 

\subsection{Silicate Feature}
\begin{figure*}
\begin{center}
\includegraphics[width=\textwidth]{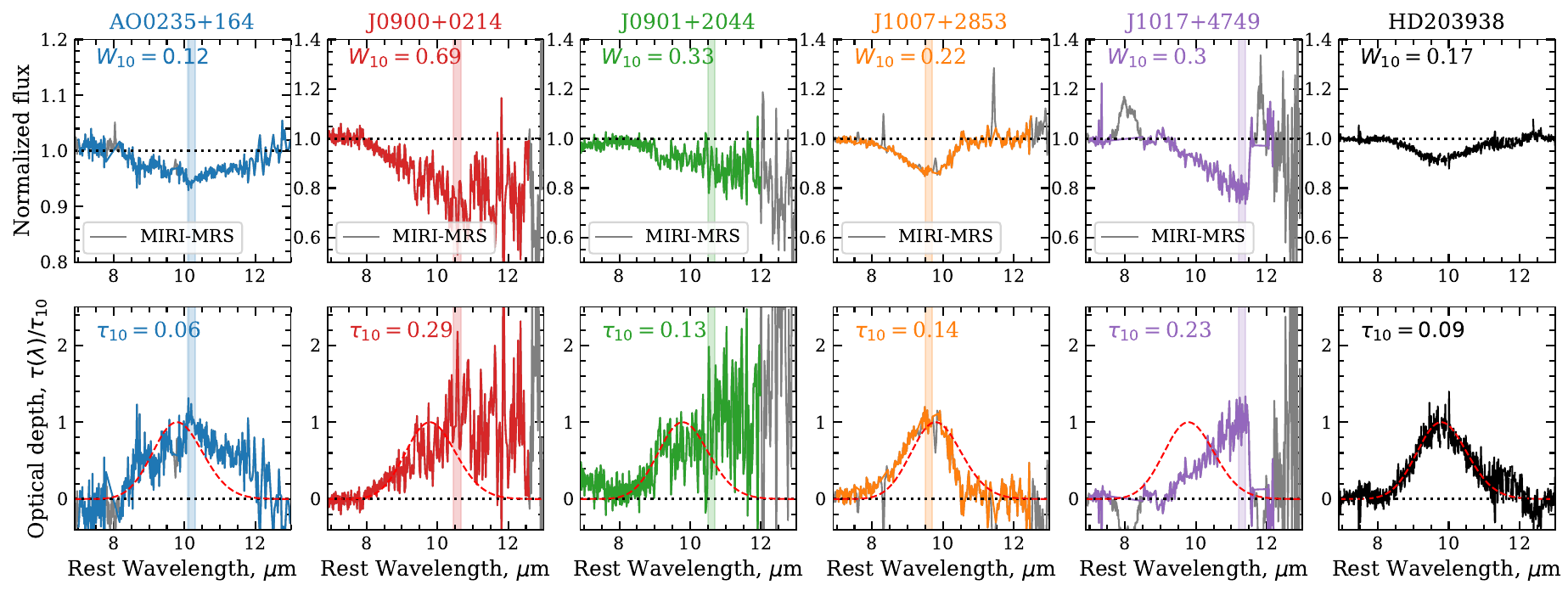}
\caption{{\it Top panel}: The 10~$\mu$m silicate
absorption in the normalized MIRI MRS quasar spectra (binned by a factor of 4) is shown in the absorber rest frame. Vertical shaded areas indicate the peak wavelengths of the observed profiles. Additionally, the rightmost panel shows the profile of the silicate absorption in the local ISM \citep[observed with MIRI MRS,][]{Decleir2025}.
{\it Bottom panels}: The optical depth profiles, normalized to the peak optical depth (or a slightly bluer peak optical depth for J0900+0214 and J0901+2044 due to the Ch4C thermal noise), calculated at the peak wavelengths (within the shaded area). The red dashed line shows the best fit to the silicate feature in the local ISM \citep{Decleir2025}.
}
\label{fig:silicate-profiles}
\end{center}
\end{figure*}

We detect the 10~$\mu$m silicate absorption feature from the foreground galaxies in all five quasar spectra. The top panels in Figure~\ref{fig:silicate-profiles} present the quasar spectra between 7-13~$\mu$m (absorber rest-frame), normalized by the AGN continua determined as described in Section~\ref{sec:continuum_model}. 
The rest equivalent widths of the silicate features ($W_{\\\rm 10}$)  were measured by integrating 
$(1 - F_{\lambda})$ in normalized quasar spectra over the range 8-12~$\mu$m of the features, and scaled  to the rest-frame. The values of
$W_{\\\rm 10}$ range from 0.1~$\mu$m to 0.7~$\mu$m, and are listed in Table~\ref{tab:results_weak}. Detections have a significance level greater than $>20\sigma$.\footnote{The uncertainties on equivalent widths include only statistical errors, and do not account for possible systematics from uncertainties in the adopted AGN continuum.}  
The bottom panels of Figure~\ref{fig:silicate-profiles} show the peak optical depth of the silicate features, calculated as $\tau=-\ln(F_{\rm norm})$,
and normalized at the optical depth near the maximum depth of the absorption feature. It is worth noting that the red wings of the silicate features for J0900+0214 and J0901+2044 overlap with the MIRI MRS channel 4C, which results in a low SNR due to higher thermal noise in channel 4C. 
Therefore, we normalize the profiles in both cases using the spectral region at bluer wavelengths. 

The profiles show a wide variety in basic properties. For example, they peak at wavelengths ranging from 9.7~$\mu$m to 11.2~$\mu$m.  
The FWHM of the features can span from 1.3~$\mu$m to 3~$\mu$m. 
The degree of asymmetry also varies, with some features showing a gentler decline on the short-wavelength side compared to the steeper long-wavelength side. 

IR dust spectral features in the Milky Way diffuse ISM ($A_V \sim 3$) have been studied in a relatively small number of sightlines, whereas numerous studies have focused on much denser and dustier regions of the 
Milky Way ($A_V \sim 10-30$) \citep[e.g.,][]{Roche1984, Roche1985, Rieke1985, Chiar2006, vanBreemen2011, Shao2018}.
Recently \cite{Gordon2021} studied silicate features in a sample 16 sightlines  ($1.8<A_V<4.6$) in archival Spitzer IRS data, and \cite{Decleir2025} in a sample of 9 sighlines ($1.2<A_V<2.5$) using the JWST NIRCam grism and MIRI MRS observations. For a basic comparison between our silicate profiles and those of the Milky Way, we chose the sightline to the star HD203938 from \cite{Decleir2025}, whose profile parameters are most similar to the average of the Milky Way sample (see Table~2 in their paper). The profile and the best fit are shown in the rightmost panels of Fig.~\ref{fig:silicate-profiles}. The data are reproduced from the supplementary materials of \cite{Decleir2025}.

Comparing profiles in distant galaxies with the profile in the HD203938 sightline, it is clear that a much wider array of 10 $\mu$m feature shapes and structures is seen in the interstellar dust in distant galaxies in our sample (e.g., AO0235+164 and J0900+0214). 
Besides the central wavelength and width of the features, the asymmetry in the features also differs from the features in the Milky Way ISM.
For example, the gentler decline on the short-wavelength side compared to the steeper long-wavelength side seen in J1007+2853 and  J1017+4749 sightlines contrasts sharply with the gentler decline on the long-wavelength side compared to the short-wavelength side seen in the silicate profile in the HD203938 sightline. 
The diversity of profiles seen in our MRS data confirm the results from our earlier studies of quasar absorber galaxies that found the peak wavelengths to range from $\sim$9.5 to $\sim$11.2 $\mu$m \citep[e.g.,][and references therein]{Aller2013,Kulkarni2016}. The differences in the peak wavelengths, widths and asymmetry of the 10 $\mu$m feature may arise from differences in grain size, shape, and composition. Uncertainties in the AGN continuum may also contribute to the observed differences (see Paper II).

\subsection{Weak IR Features}
We also search for weak IR absorption at 3.4~$\mu$m and 6.2~$\mu$m from 
carbonaceous grains and for absorption at 3.0~$\mu$m from water ice.
These features are commonly detected in dusty regions of the ISM with high $A_V \ge 10$ \citep{Sandford1991, Chiar2013, Hensley2020}, and were recently observed in the diffuse ISM at lower extinctions ($A_V \leq 2.5$; \citealt{Decleir2025}).

In the Milky Way diffuse ISM, the strengths of the 3.4~$\mu$m and 6.2~$\mu$m features are $A_V / \tau_{\rm 3.4} \sim 280$ and $A_V / \tau_{\rm 6.2} \sim 310$, respectively, which roughly correspond to ratios of $\tau_{\rm 3.4}/\tau_{10}$ and $\tau_{\rm 6.2}/\tau_{10}$ of about $\sim0.1$ \citep[see][]{Decleir2025}. Similar values are also predicted in chemical models \citep{Gao2010}.
Considering the measured values of $\tau_{\rm 10}$ in our sample, we can estimate the strengths of the 3.4~$\mu$m and 6.2~$\mu$m absorptions to be about $\tau\sim0.01-0.03$, or 1-3\% of the normalized flux. These features would be difficult to detect given the SNR of our data. To improve the SNR, we rebinned our data by a factor of 10.

\begin{figure*}
\begin{center}
\includegraphics[width=\textwidth]{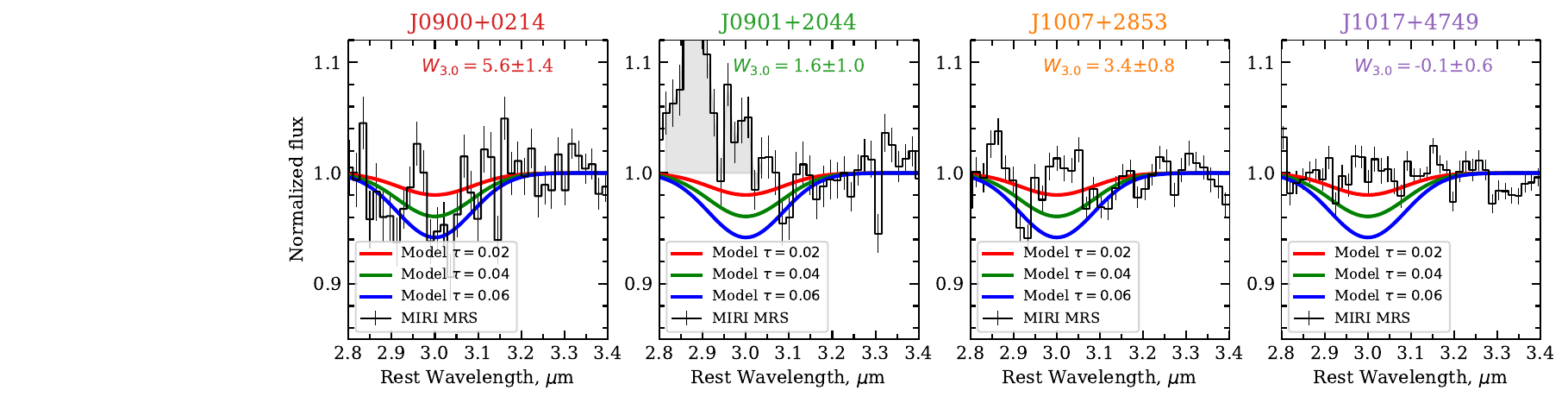}
\includegraphics[width=\textwidth]{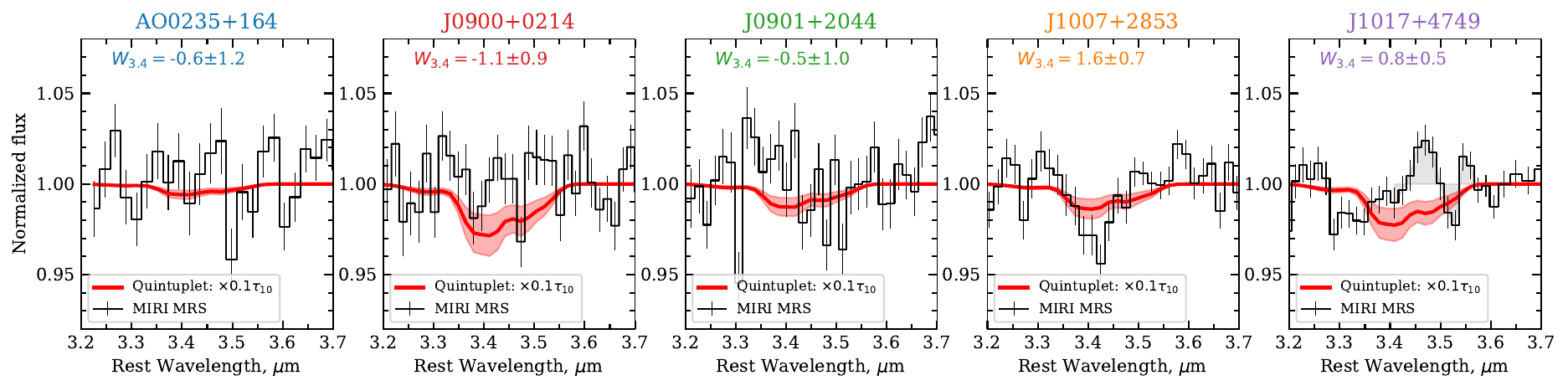}
\includegraphics[width=\textwidth]{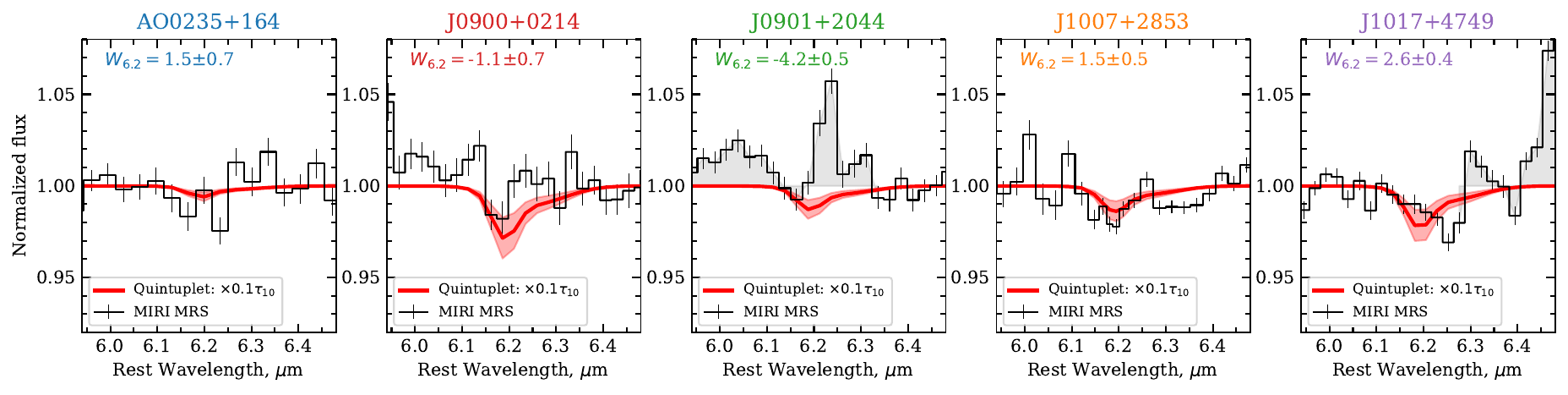}
\caption{Portions of the normalized MIRI MRS quasar spectra (binned by a factor of 10) showing the 3~$\mu$m (top panels), 3.4~$\mu$m (middle panels) and 6.2~$\mu$m (bottom panels) ice/dust feature locations in the rest frame of the foreground, absorber galaxies. The equivalent width of features (in $10^{-3}~\mu$m) is shown in the upper left corner of the panels. For the 3~$\mu$m feature, the red, green and blue lines show examples of the profile of the feature, centered at 3.0~$\mu$m, and normalized at the peak optical depth of $\tau_3=0.02$, 0.04 and 0.06, respectively. For the 3.4~$\mu$m and 6.2~$\mu$m features, the red line shows the profile for the Galactic center Quintuplet Cluster as fit following \cite{Chiar2013}, assuming an optical depth value of $\tau=0.1\tau_{10}$. The red shaded area shows the uncertainty of the profile due to the range of possible values of $\tau/\tau_{10}=0.06-0.14$, as it suggested by \cite{Gao2010}. The gray shaded area indicates regions overlapping with quasar emission lines. 
}
\label{fig:weak-profiles}
\end{center}
\end{figure*}

Figure~\ref{fig:weak-profiles} compares our rebinned data and synthetic profiles of the 3.4~$\mu$m and 6.2~$\mu$m features, calculated using a model from \cite{Chiar2013} and assuming optical depth values of  0.1 $\tau_{10}$. The absorptions are weak, and likely lie within the statistical noise of the data.  In some cases, the features additionally overlap with quasar emission lines (3.4~$\mu$m for J1017+4749; 6.2~$\mu$m for J0901$+$2044 and  J1017+4749), which further complicates their detections. However, one case, J1007+2853, shows a tentative detection of the 3.4~$\mu$m feature. We can quantify this by measuring the equivalent width, which is greater than zero, at a significance level of  $\sim$ 2.3$\sigma$  (see Figure~\ref{fig:weak-profiles} and  Table~\ref{tab:results_weak}); this is a little stronger than predicted by assuming the $\tau_{34}/\tau_{10}$ ratio to be 0.1. Follow-up, higher-SNR observations are needed to confirm this detection.

H$_2$O ice features are not commonly detected in diffuse, low-$A_V$ sightlines. However, the 3~$\mu$m feature was recently detected by \cite{Decleir2025} in one of nine diffuse sightlines observed with the JWST. The top panels in Fig.~\ref{fig:weak-profiles} show regions near 3~$\mu$m in the absorber rest-frame for our sample sources. For comparison, we show Gaussian profiles, centered at 3.0~$\mu$m, for optical depths of $\tau_{3.0}=0.02-0.06$. While the data are noisy, in the case of J0900+0214 we detect a possible absorption feature at 3.0~$\mu$m with an equivalent width above zero at the 2.5$\sigma$ level.

\setlength{\tabcolsep}{2pt}
\begin{table}
\begin{center}
\caption{The rest equivalent widths of the IR dust features  in our sample.}
\label{tab:results_weak}
\begin{tabular}{|l|c|c|c|c|c|}
\hline
  Quasar & $A_V$ & $W_{3}$ & $W_{3.4}$ & $W_{6.2}$ & $W_{10}$\\
   &  mag & 10$^{-3}\mu$m  & 10$^{-3}\mu$m  &  10$^{-3}\mu$m & 10$^{-3}\mu$m   \\
\hline
 AO0235+164 & $\sim0.5$ & .. & $<3$ & $<2$  & $123\pm2$ \\ 
 J0900+0214 & ..    & $5\pm2$ & $<3$ &  $<2$ & $690\pm25$ \\ 
 J0901+2044 & $0.22^{+0.40}_{-0.20}$ & $2\pm1$ & $<3$ & ..  & $330\pm12$ \\ 
 J1007+2853 &  $1.08^{+0.11}_{-0.11}$ & $<3$ & $2\pm1$ & $<3$ & $220\pm2$  \\ 
 J1017+4749 & .. &  $<3$ &  $<2$  & $<4$ & $296\pm4$  \\ 
\hline
\end{tabular}
\begin{tablenotes}
      \small
     \item{Notes: The rest equivalent widths are measured in units of $10^{-3}~\mu$m for features  at 3~$\mu$m (H$_2$O), 3.4~$\mu$m (C–H aliphatic hydrocarbon), 6.2~$\mu$m (C=C olefenic/aromatic hydrocarbon), and 10~$\mu$m (silicate).}
    \end{tablenotes}
\end{center}
\end{table}

\section{Conclusions}
\label{sec:conclusions}
This analysis is the first paper of our study of dust features in distant galaxies using high spectral resolution observations with the JWST MIRI MRS. The project aims to investigate evidence of dust evolution in distant galaxies, and to verify previous findings from our team, based on Spitzer IRS spectroscopy, that silicate dust grains show a greater diversity in properties at high-redshift than seen in the Milky Way.

We have performed the first high spectral resolution observations of infrared dust features arising from intervening galaxies along sightlines to higher-redshift, background quasars, using the JWST MIRI MRS.
The five quasars were chosen to have strong Mg~{\sc ii} absorbers at $0.5 < z < 1.1$, and to show a significant 2175 \AA\ bump dust feature in absorption.
In all sightlines, we have detected and measured silicate 10~$\mu$m absorption. 

The high resolution of the MIRI MRS allows us to accurately measure the strength and profile shape of this 10 $\mu$m silicate absorption, and to tentatively detect weak water ice and dust features at 3~$\mu$m and 3.4~$\mu$m. 
Our main results are as follows:

(1)  The MIRI MRS is a powerful instrument for studying weak infrared dust features originating from intervening galaxies in quasar spectra. However, because distant quasars are usually observed as faint sources, with fluxes of several mJy, careful attention must be given to cleaning the observational data of instrumental artifacts, such as scattered light and cosmic ray showers. These artifacts can easily mimic or enhance broad and weak infrared dust features in the quasar spectra. We present a procedure for identifying artifacts and removing them from both background and target exposures.

(2) One of our targets, AO0235+164 has been previously observed by our team with the Spitzer IRS. While this quasar is a temporally-variable blazar, we found a good consistency between the 10~$\mu$m silicate feature profile in normalized spectra observed in 2006 with the Spitzer IRS and in 2023 with the MIRI MRS. This demonstrates the accurate flux calibration of the MIRI MRS and the robustness of our reduction procedures.  

(3) The 10~$\mu$m silicate feature has been detected in all of our targets. This demonstrates that quasar absorbers with strong 2175 \AA\ bump strengths are good candidates for studying dust silicate features in the diffuse ISM of distant galaxies. Furthermore, systems which are rich in carbonaceous grains may also exhibit significant quantities of silicate dust.

(4) The profiles of the 10~$\mu$m silicate feature show a wide diversity of peak wavelengths and shapes among our sample sources, as well as differences relative to the average profile in the Milky Way diffuse ISM. The peak wavelength of our features ranges from 9.7~$\mu$m to 11.2~$\mu$m. Some features show broader or narrower profiles compared with the average profile in the Milky Way ISM. The higher-redshift features also demonstrate varying degrees of asymmetry, 
showing both more symmetric or asymmetric profiles. These variations are likely indicative of differences in the grain mineralogy and structure, which will be discussed further in Paper II.

(5) We have tentatively detected weak infrared features at 3.0~$\mu$m (for J0900+0214) and 3.4~$\mu$m (for J1007+2853), likely caused by H$_2$O-ice and by aliphatic hydrocarbons (C-H mode), respectively. While these data are noisy, equivalent widths are above zero at the 2.5$\sigma$ and 2.3$\sigma$ levels, respectively. Follow-up observations are needed to confirm this detection. 

\section{Data availability}

Data directly related to this publication and its figures can be requested from the authors. The {\it JWST} data used in this paper are publicly available, and can be found in the MAST (Mikulski Archive for Space Telescopes) portal under program ID 2155,  DOI: 10.17909/br31-0120. 

\section*{Acknowledgements}
We thank an anonymous referee for constructive comments that have helped to improve this paper. We would like to thank Marjorie Decleir for providing the data ahead of publication.
This work is supported by a grant from the Space Telescope Science Institute for JWST GO program 2155 (PI V. Kulkarni). Additional support from NASA grants 80NSSC20K0887 and 80NSSC24K1162 (PI V. Kulkarni) is also gratefully acknowledged. This work is based in part on observations made with the NASA/ESA/CSA James Webb Space Telescope. The data were downloaded from the Mikulski Archive for Space Telescopes at the Space Telescope Science Institute which is operated by the Association of Universities for Research in Astronomy, Inc., under NASA contract NAS 5-03127 for JWST. Our observations are associated with program 2155. Support for program 2155 was provided by NASA through a grant from the Space Telescope Science Institute, which is operated by the Association of Universities for Research in Astronomy, Inc., under NASA contract NAS 5-03127. 

\bibliographystyle{mnras}
\bibliography{Library} 

\newpage
\appendix

\section{Details of MIRI MRS reduction}
\label{app:A}
In this Appendix, we provide a more detailed description of the custom procedures applied to our MIRI MRS data. We aim to demonstrate the data improvements achieved through our procedure for cosmic-ray (CR) shower corrections in background fields and target exposures.\footnote{The most recent version (11.3) of the pipeline has improved corrections for cosmic ray showers in MRS data%
(\url{https://jwst-docs.stsci.edu/known-issues-with-jwst-data/shower-and-snowball-artifacts}); 
however, it was not available at the time of our data reduction.} Additionally, we highlight a new artifact, potentially caused by switching the MIRI MRS readout mode from the SLOW to the FAST readout pattern. This artifact was identified in the observations for 2 out of our 5 targets. 

\subsection{Correction for CR showers in background exposures}
\label{app:A:Background-model}

\begin{figure*}
\begin{center}
        \includegraphics[width=\textwidth]{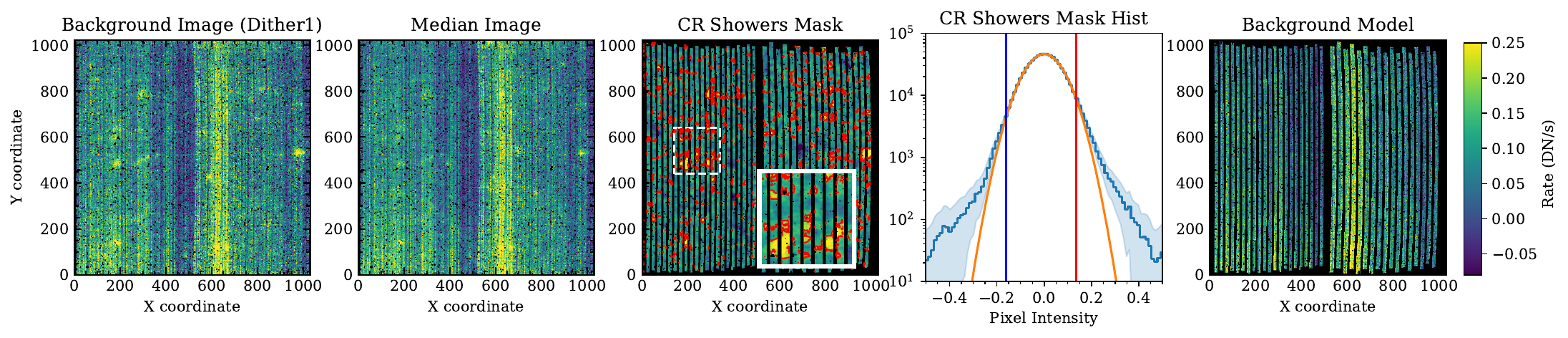}
        \includegraphics[width=\textwidth]{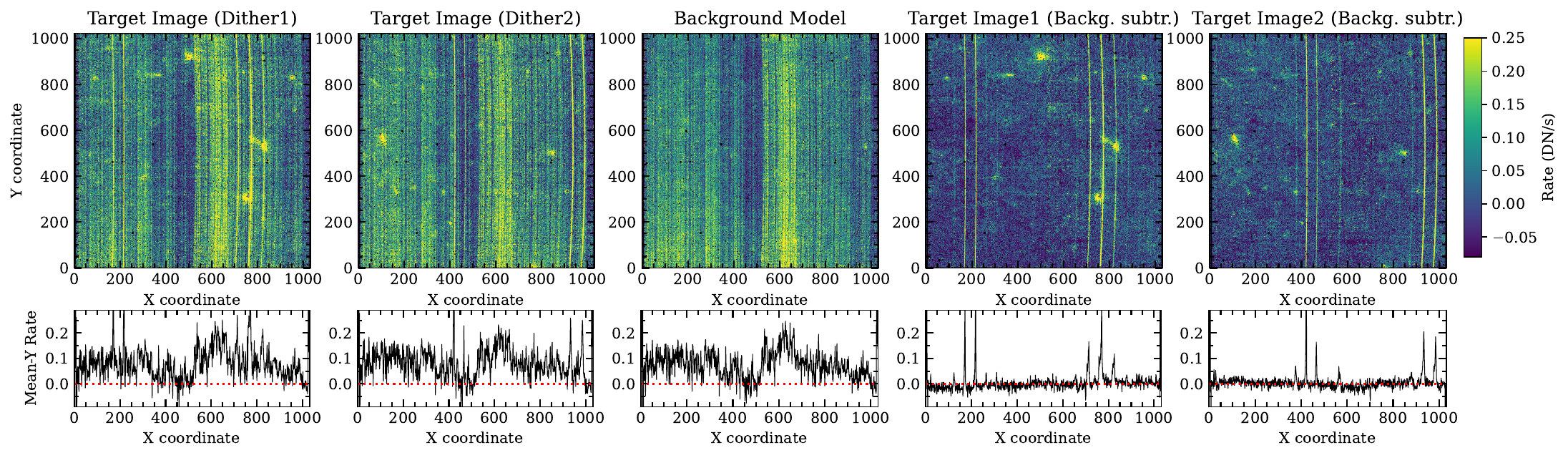}
        \caption{An illustration of the procedure to correct for CR showers in the background exposures. 
        The {\it top panels} present, from left to right: (i) the MIRI MRS detector image of the background field for the QSO J0901+2044, showing data for channels 1,2 and medium band (MIRIFUSHORT); (ii) the median image of the dithered background images; (iii) the cosmic ray (CR)-showers mask; (iv) the histogram of the pixel rate in the CR showers mask; and (v) the background model. The color in panels (i), (ii) and (v) and the color bar encodes the pixel rate in DN/s. The color in panel (iii) encodes the relative difference between panels (i) and (ii) and ranges between -0.3 and 0.3 units.
        The black pixels in panels (iii) and (v) are located between the IFU slices, and do not contain data from the observations.
        The small panel in the bottom right corner shows a zoomed-in view of the region marked by the white dashed rectangle. The red contours in the panel (iii) indicate the pixels affected by CR showers. They are identified as positive outliers in the histogram shown in panel (iv); see text for details. 
        The blue and orange curves in panel (iv) denote the histogram and the best fit with a Gaussian profile, respectively. The blue shaded area indicates the $\pm10\sigma$  uncertainty in the number of counts. The red and blue vertical lines show the threshold values where the histogram deviates from the Gaussian fit by 10$\sigma$. The {\it bottom panels} present, from left to right: (i,ii) - the MIRIFUSHORT images for the quasar J0901+2044 field with dithers 1 and 2, respectively; (iii) the background model, as in top panel (v), with pixels between the IFU slices not masked for consistency with the presentation of the other panels; (iv, v) - the MIRI MRS target images with the background subtracted. The color gradient in all panels has the same scale, and represents the pixel rate in DN/s. The bottom sub-panels show the corresponding profiles of the pixel rate averaged along the dispersion (Y) axis.}
        \label{fig:CR-correction}
\end{center}
\end{figure*}

Here, we describe our methodology to correct the images for CR showers (residual effects of energetic CR hits) that cause a slight increase (about $\sim$10\%) in the brightness of nearby pixels due to inter-pixel charge migration, and appear as localized bright spots in both the background and in the target images. This effect is illustrated in Figure~\ref{fig:CR-correction}, which shows example MIRIFUSHORT detector images for background ('Dither 1' in the top panels) and target ('Dither 1' and 'Dither 2' in the bottom panels) exposures taken for the quasar J0901+2044 in the medium band. The detector image simultaneously contains data for two channels: 1 and 2 (the left and right parts of the detector).\footnote{MRS spectra from all four channels are dispersed simultaneously onto two detectors (MIRIFUSHORT and MIRIFULONG) for a single exposure \cite[e.g.,][]{Argyriou2023B}.}
The vertical traces in multiple slices along the dispersion axis in the target images represent the quasar emission. The background pattern is similar for both the background and target exposures.

Assuming that the background sky fields do not contain unresolved sources, we expect that corresponding pixels across the dithered background images\footnote{In our program, background fields were observed with either 2 or 4 dithers (see Table~\ref{tab:obslog}).} should have similar brightness, varying only due to statistical noise. Therefore, we can combine the dithered background images to improve the signal to noise of the background model. The panel 'Median Image' in  Figure~\ref{fig:CR-correction} shows the median image for the dithered background images (two background images were observed for J0901+2044). The intensity of the CR showers decreases by a factor of two; however, they still remain in the median background image.

At pixel-level resolution, statistical noise typically dominates over distortions caused by CR showers, making it difficult to directly detect affected pixels through pixel-to-pixel intensity comparisons. Our procedure takes advantage of the spatial correlation among affected pixels, which makes them easier to detect in smoothed detector images. This is an iterative process.
In the first iteration, the background model is the median of the dithered background images. For the background model, we calculate ``Masks'' for the background images as the relative difference between the dithered image and the background model (i.e., $Mask=Image/Model - 1$). Then, we first exclude pixels that lie between the MRS IFU slices. These pixels are not illuminated during observations, and show a strong difference in intensity, compared with the illuminated pixels in the IFU slices, especially in images for the 3/4-channels (in MIRIFULONG detector), which may affect the subsequent smoothing step. After that, we smooth the ``Masks'' using a circular filter with a 10-pixel radius. The smoothed ``Mask'' image is shown in the panel 'CR Showers Mask' of Figure~\ref{fig:CR-correction}. The inter-slice pixels are shown in black, and the color gradient corresponds to $-0.3$ to 0.3 units.

The pixels affected by CR showers in the Dither~1 Image become more visible, and appear as 'yellow' positive spots, while most of the pixels (`normal' pixels) have their intensity near zero. We identify the affected pixels by setting a threshold on the brightness of the outliers.
To do this, we consider the distribution of pixel brightness for the entire ``Mask'' image. This histogram is shown in the `CR Showers Mask Hist' panel. The intensity of normal pixels are well-described by a Gaussian distribution. The outliers with positive (yellow) and negative (blue) intensity correspond to pixels affected by CR showers in the Dither 1 and 2 images, respectively\footnote{This is  correct if the background field was observed with two exposures. For a larger number of dithered exposure, the median image is less affected by CR showers in individual exposures, providing a cleaner reference. In that case, the CR Shower Mask image will show only positive excess (in yellow) of the flux due to CR showers in the associated dithered image. The negative excess (in blue) is possible, but it indicates instrumental artifacts in the analyzed dithered image, whereas in the case of dither = 2, it indicates CR showers in the second dithered image.}. Given that the statistical uncertainty in the number of counts per bin is the square root of the number of counts, we define the threshold as the value where the difference between the histogram and the Gaussian model exceeds the 10$\sigma$ level. The positive outliers are highlighted in red contours in the `CR Showers Mask' panel. Their position correlates with the position of CR showers in the `Background Image  (Dither 1)'.
We tested the procedure on mock data, and found good consistency between the selected outliers and the simulated CR showers.

In the second iteration, the background model is calculated again as the median of the original exposures. However, we correct these exposures for two effects: (i) we exclude the positive outliers in the `CR Showers Mask' identified in the first iteration, and (ii) we found that the distribution of pixel brightness in the `CR Showers Mask' is not always centered at zero. This happens due to a variability in the average brightness of the background images. In order to correct for this, we subtract the background model (calculated at the previous iteration), scaled by the offset value, from the original background image. 
Then, using a new background model, we search for CR showers and determine the new offset. Usually three iterations are sufficient for a realistic model of the background image.

The bottom panels in Figure~\ref{fig:CR-correction} show the target images before and after subtraction of the background model.  The background model is the same as shown in the upper panels. For clarity of illustration, we also show the intensity in the masked pixels (between the IFU slices), calculated as the median of the corresponding pixels in the dithered images. These pixels do not contain important information, as they are masked later (set as NaN) during the flux calibration step. It is worth pointing out that our quasars are very faint, with intensity levels just slightly above the MIRI MRS background. That makes correcting for background variations especially important for studying weak features in their spectra.
In the bottom sub-panels, we show the profiles of the averaged intensity along the dispersion (Y) axis. We demonstrate that for background-subtracted images, the average intensity outside the quasar traces is consistent with the zero level, within a few percent. 

At the same time, it is clear that the background-subtracted images exhibit their own CR showers, which also need to be corrected. Unfortunately, we cannot apply the same procedure as for the background images, because dithering shifts the quasar traces and potentially features related to other objects in the target fields, such as absorber host galaxies. A direct pixel-by-pixel comparison in the detector-plane coordinate system will flag these real features as outliers. Even if we mask the pixels along the quasar traces, it will not help remove CR showers that overlap with the quasar traces. To correct data for CR showers in the target background-subtracted exposures, we developed a procedure which is described in Section~\ref{app:A:res_bkgr_corr}. 

\subsection{Correction for CR Showers in Target Exposures}
\label{app:A:res_bkgr_corr}

As shown in the previous section, the background subtracted images contain quasar traces (and may contain features of the absorbers' host galaxies) and artifacts  - CR showers, while the background of these images is consistent with the zero level. Some of the CR showers overlap with the quasar traces, increasing the count rate in the affected pixels. 
Due to dithering, the pixels related to quasar emission change their coordinates in the detector coordinates plane. However, these pixels remain in the same positions in sky coordinates. Direct comparison of pixels with the same sky coordinates in detector images is complicated, but it becomes easier when working with cube data.

By default, the pipeline makes a correction for CR showers by averaging data between dithered detector images while creating a combined IFU cube. However, we aim to make a more precise correction, by constructing a model for the CR showers in the cube data separately for each dithered image, and removing them from data before extracting the final quasar spectrum.

\begin{figure*}
\begin{center}
        \includegraphics[width=0.9\textwidth]{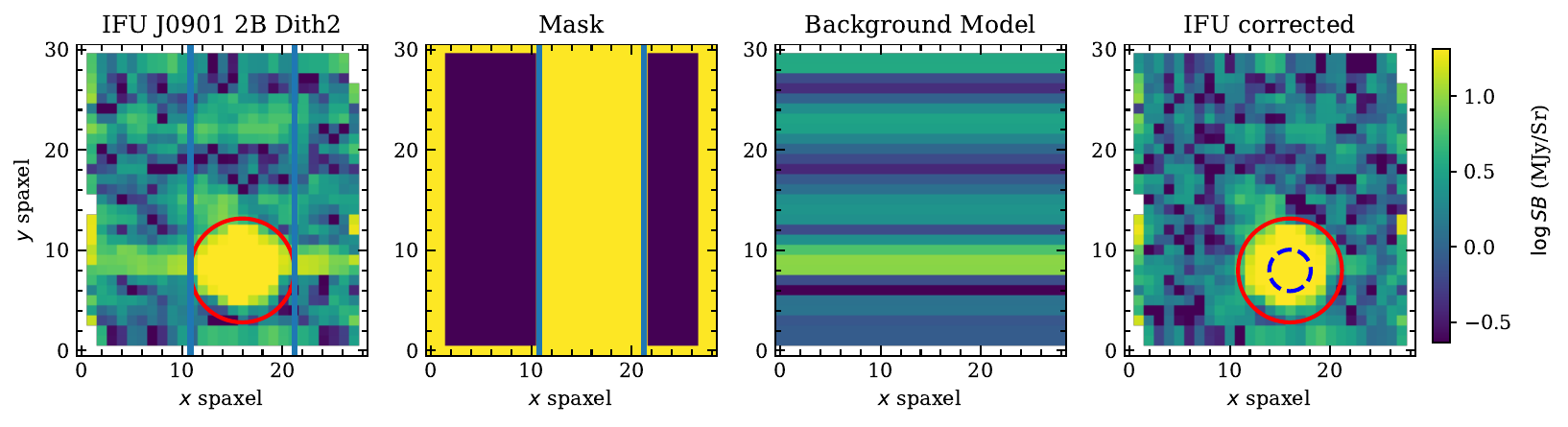}
        \includegraphics[width=0.9\textwidth]{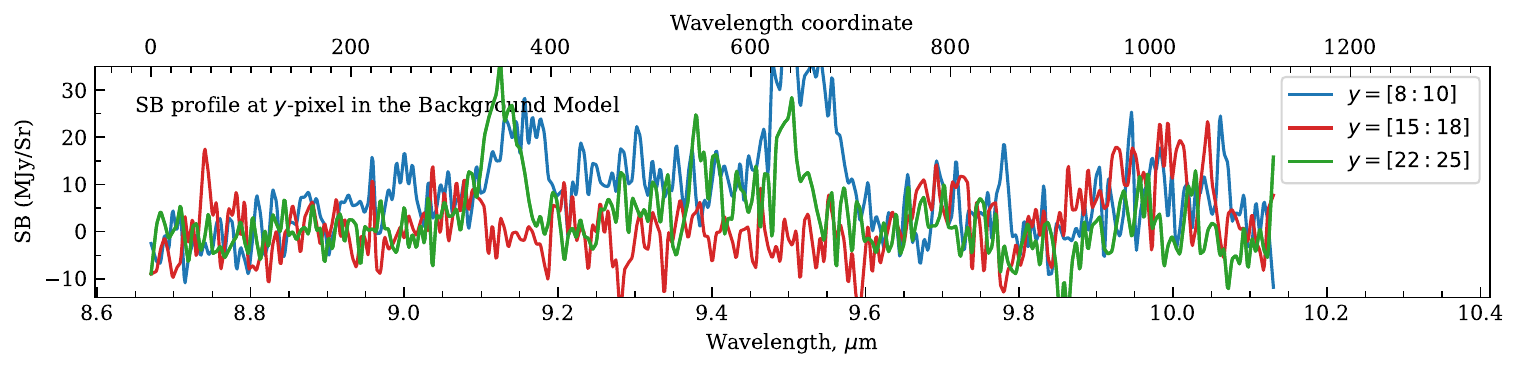}
        \includegraphics[width=0.9\textwidth]{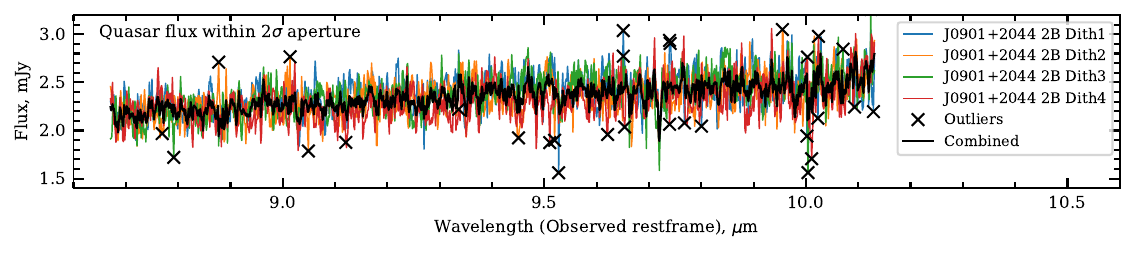}
        \caption{
        An illustration of the procedure for correction for CR showers in the target exposures. As an example, we show data for the quasar J0901+2044 in the MIRI channel 2B (Dither 2). The {\it top panels} show, from left to right: (i) an image of the IFU cube, averaged along the dispersion axis and shown in the coordinate system, aligned with the instrument IFU plane; (ii) the mask for spaxels used to construct the background model; (iii) an image of the background model; and (iv) an image of the IFU cube after subtraction of the background model.
        The color gradient in panels (i), (iii), (iv), and in the color bar, encodes the pixels surface brightness (SB) in MJy/Sr.
        The red and blue circles indicate the area of the $5\sigma$ and $2\sigma$ widths of the point spread function (PSF) around the quasar. Blue vertical lines denote a shift of $\pm$5$\sigma$ from the quasar position along the horizontal axis.  
        The {\it middle panel} shows the spectral SB profiles for spaxels in the background model with $y$-coordinates in three ranges: [8:10] (blue), [15:18] (red), and [22:25] (green). The positions of the SB spikes at $\sim9.1~\mu$m and $\sim9.5~\mu$m correspond to the $Y$-coordinates of CR showers in the detector image (yellow spots in the bottom rightmost panel in Fig.~\ref{fig:CR-correction} in the right part of the detector image at $Y\simeq300$ and 550). 
        The {\it bottom panel} compares four spectra of the quasar, extracted with an aperture of the 2$\sigma$ width of the MIRI MRS PSF from the background-corrected IFU cubes with dither positions from 1 to 4. 
        The spectra show strong consistency. The black curve represents the combined quasar spectrum. The $3\sigma$ outliers are marked by black crosses.}
        \label{fig:app:cube-corrections}
\end{center}
\end{figure*}

For each target detector image with the background subtracted (with dither positions 1 to 4), we created an IFU cube using a coordinate system aligned with the instrument IFU plane. This alignment ensures that the spaxels along the horizontal axis (in IFU cube) correspond\footnote{A schematic overview of the MIRI MRS IFU is shown in Fig.~2 of \cite{Argyriou2023B}. The IFU cube/image is constructed from several horizontal slices that share the same horizontal ($x$) coordinates but differ in vertical ($y$) coordinates in the IFU-aligned coordinate system. The $y$-coordinate roughly corresponds to the slice number ($N_i$), indicating the slice's position in the detector image. In the detector image, these slices appear as narrow vertical stripes. The vertical ($Y$) coordinate corresponds to the dispersion axis ($\lambda$), and the slice number ($N_i$) determines the horizontal ($X$) offset of the stripe on the detector. For pixels within a single stripe, the horizontal ($X$) coordinate, corrected for the offset, corresponds to the $x$ coordinate of the spaxels in the IFU cube/image.
}
to the pixels in the detector images along the X-coordinate within each (vertical) slice. An example of a cube image (integrated along the dispersion axis) is shown in the upper left panel of Figure~\ref{fig:app:cube-corrections}. 
Due to the presence of CR showers in the detector image (see panel (v) in the lower panels of Figure~\ref{fig:CR-correction}), the background pixels around the quasar in the cube image exhibit some artifacts. Below we describe the procedure which corrects for such artifacts in cube data. 

The general idea is to create a model for the background around the quasar using pixels outside of the quasar's PSF, and to subtract this model from the data.
First, we mask the spaxels within 5$\sigma$ of the MIRI MRS PSF function\footnote{The MRS empirical PSFs were taken from \cite{Argyriou2023B}.} around the quasar. We also masked spaxels at the edge of IFU, to exclude the flux variation near the edge of the detector. We note that in the IFU-aligned coordinate system, artifacts due to CR showers appear as features aligned along the horizontal axis\footnote{This occurs because the IFU is constructed from horizontal slices, which appear in the detector image as narrow vertical stripes. The width of these stripes is approximately 25 pixels, comparable to the typical size of a cosmic ray (CR) shower. Therefore, a stripe affected by an energetic CR shower will exhibit excess brightness (at specific wavelengths and in the average), compared to nearby slices. Such a slice will appear in the IFU cube (IFU image) as a horizontal line artifact over a certain wavelength range (in the average brightness).
}. If a CR shower overlaps with the quasar trace, we can model its intensity in the quasar spaxels using the spaxels to the left and right of the quasar 5$\sigma$ area.  An example of the mask is shown in the second top panel in Figure~\ref{fig:app:cube-corrections}. The unmasked area above the quasar is used to test the consistency of background subtracted data with zero-flux. 
We note that some quasar fields may have several sources (e.g., absorber host galaxies around AO0235+164), which we further mask to prevent contamination of the background model.

Using this mask, we construct a background model by averaging spectra in unmasked spaxels along the horizontal axis. Therefore, the pixel intensity in the background model depends on two coordinates: one along the vertical axis (that corresponds to different slices in a detector image in Figure~\ref{fig:CR-correction}) and the other along the dispersion axis (the vertical axis in a detector image). 
The mean SB image of the background model is shown in the third top panel from left in Figure~\ref{fig:app:cube-corrections}. The middle panel shows the SB profile of several spaxels in the background model for the range of $y$ coordinates, as indicated in the legend. Note that the $y$-coordinate corresponds to the position (along the $X$ axis) of the stripe in the detector image. We depict several spikes with blue and green spectra (at 9.1 and 9.5~$\mu$m or $\sim350$ and $\sim600$ units in the dispersion axis, respectively), which originate from CR showers seen in the detector image at the $Y$ coordinate of $\simeq$300 and 550 units along the slices containing quasar traces (see the bottom rightmost panel in Figure~\ref{fig:CR-correction}). Note that there is a one-to-one correspondence between the $Y$ coordinate in the detector image and the pixel number in the dispersion array of the cube, due to the bending of the stripes in the detector images (and the Y offset).

After subtracting the background model, the SB in the field around the quasar is well-consistent with the zero-level (see the top rightmost panel in Figure~\ref{fig:app:cube-corrections}). Next, we
extract the quasar spectrum from spaxels within a 2$\sigma$ aperture around the quasar (shown by the blue dashed circle). The bottom panel in Figure~\ref{fig:app:cube-corrections} shows four quasar spectra, extracted from the background-subtracted cubes for four dither positions. The spectra match each other well and do not show features related to CR showers. The combined mean spectrum, shown in black, closely matches the fluxes from the individual exposures. We also identify potential outliers using a 3$\sigma$ clipping method to ensure accurate data.

\subsection{Artifact in MIRI MRS data}
\label{app:A:artifact}

We report an artifact identified in the MIRIFULONG\footnote{MIRIFULONG detector contains data from channels 3 and 4.} detector data, observed exclusively in two of the five quasars: AO0235+164 and J0901+2044, affecting data in the SHORT and LONG bands, respectively\footnote{For J0901+2044 this artifact presence in the LONG band of the channel 3/4 images, but not in  SHORT and MEDIUM bands. However, for AO0235+164, the artifact appears in the channel 3/4-detector images across all three bands, with the most prominent presence in the SHORT band.}. A large-scale background variation resembling scattered light is observed across the detector image, with a changing angle and a decrease in flux toward the bottom-left corner. This artifact is observed in all four dithered images, with its intensity gradually diminishing as the dither number increases, indicating a relaxation over time. 

\begin{figure*}
\begin{center}
        \includegraphics[width=\textwidth]{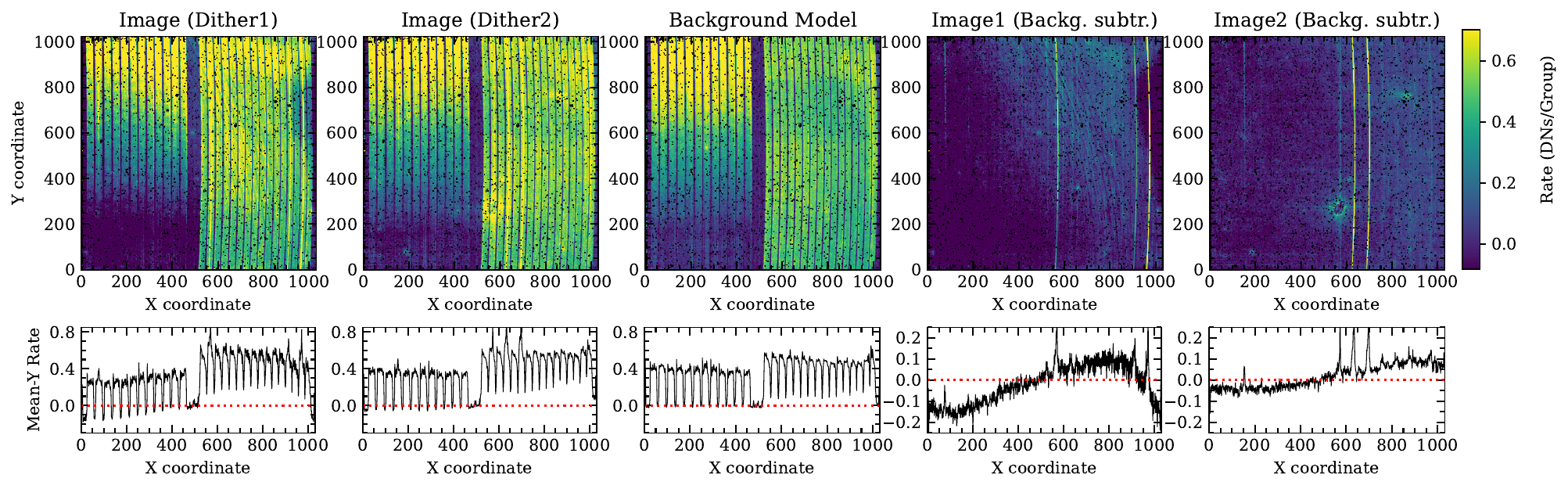}
        \includegraphics[width=0.8\textwidth]{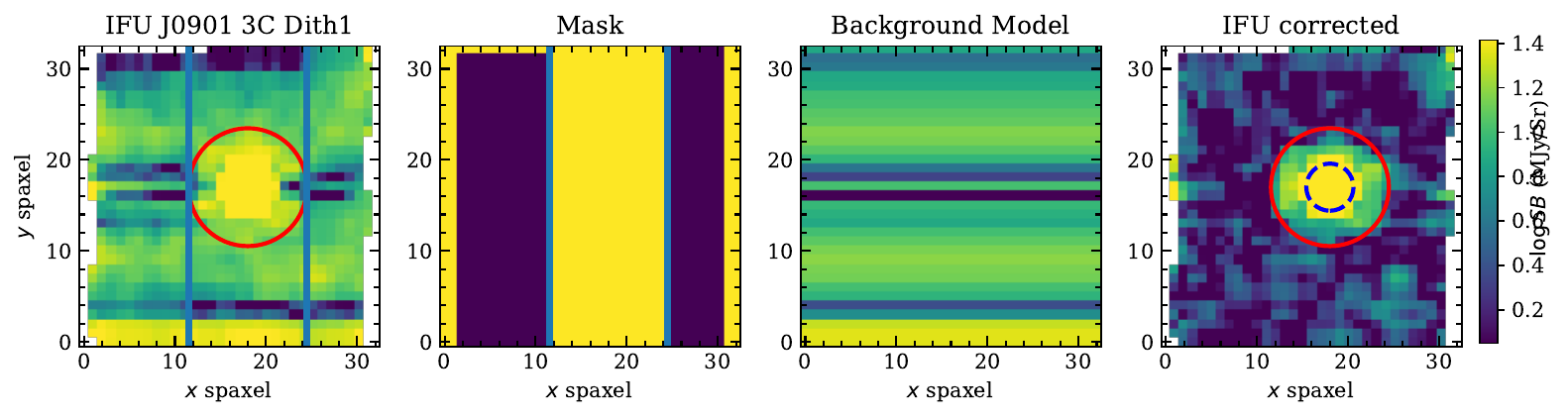}
        \includegraphics[width=0.9\textwidth]{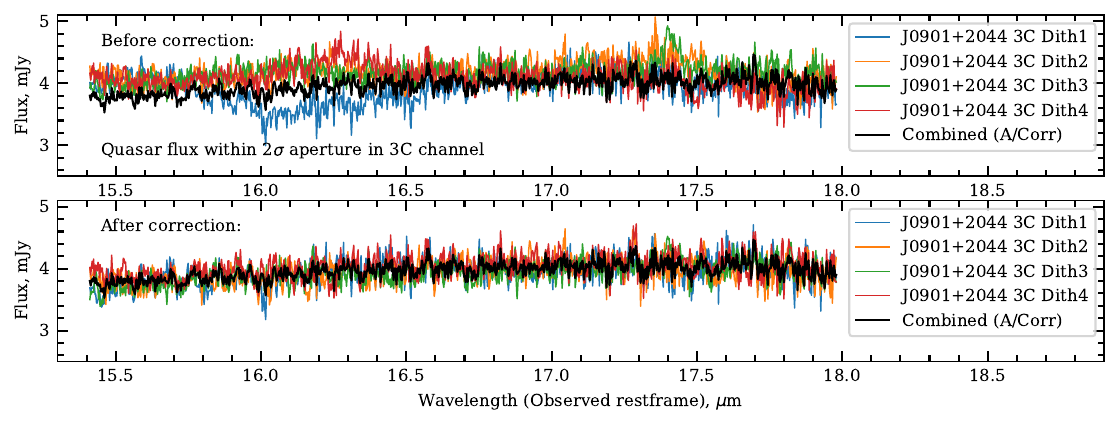}
        \caption{The figure demonstrates the artifact found in MIRI MRS exposures of the quasar J0901+2044 in channels 3/4 (MIRIFULONG) and the LONG band.
        The {\it top panels} show, from left to right: (i and ii) target detector images for dither positions 1 and 2, respectively; (iii) the background model; (iv and v) the detector images with the background subtracted. The bottom sub-panels show the rate averaged along the dispersion (Y) axis.  
        A large-scale background variation resembling scattered light is observed across the detector image, with a changing angle and a decrease in flux toward the bottom-left corner. The artifact is more 
        pronounced in the dither 1 image. 
        The {\it middle panels} show the procedure for correcting the artifact contributions in the IFU cube for the channel 3C and Dither 1. The descriptions of the panels are the same as for the top panels in Figure~\ref{fig:app:cube-corrections}. 
        The {\it bottom panels}
        show the quasar flux extracted within a 2$\sigma$ aperture using IFU data with dither positions 1 to 4, before and after the corrections. The black curve in both panels denotes the combined quasar spectrum for the data after the background corrections. The variation of the quasar spectra between dithers 1 through 4, extracted before the correction was applied, demonstrates the effect of the artifact on the quasar flux within a 2$\sigma$ aperture.
        }
        \label{fig:app:artifact-detector}
\end{center}
\end{figure*}

Fig.~\ref{fig:app:artifact-detector} presents an example of this artifact for the quasar J0901+2044 for the MIRIFULONG detector (channels 3,4) and LONG band. In the first row, we show the images for dither positions 1 and 2, where the artifact is more pronounced. We note that this artifact is not present in the background exposures. The images after background subtraction are shown in the fourth and fifth panels from the left. For the Dither 1 image, one can see dark blue (`negative' rate) spots near the bottom left corner and the top right corner. There is also a periodic pattern of `excess light' between these two `negative' spots. The bottom sub-panels (in the first row) show the pixel rate averaged along the dispersion  (Y) axis, and highlight the difference of the residual background from the zero-flux level with an amplitude of distortions of $\sim0.15$~DN/s and $\sim0.1$~DN/s for the the dither 1 and dither 2 images, respectively.   

We consulted with the STScI JWST Help Desk Team, and they confirmed that a similar artifact in data is observed just after a switch between FASTR1/SLOWR1 readout mode, and only for the MIRIFULONG detector. This can be due to a small change in the bias across the detector during the switch. To the best of our knowledge, the pipeline does not have a procedure to correct for this artifact.

We found that our procedure for correcting for CR showers in cube data is effective in removing this artifact as well. The middle and bottom panels of Fig.~\ref{fig:app:artifact-detector} illustrate the improvements for the channel 3C. Before the correction, quasar spectra extracted within a 2$\sigma$ aperture show flux variations with amplitudes of 0.1-0.2~mJy (due to scattered light) and an absorption feature at 16.2$\mu$m in the Dither 1 spectrum (caused by a dark blue spot in the top right corner). After the correction, quasar spectra from Dithers 1-4 are free of such features and show good agreement.

\end{document}